\documentclass[10pt,journal,compsoc]{IEEEtran}
\ifCLASSOPTIONcompsoc
  \usepackage[nocompress]{cite}
\else
  \usepackage{cite}
\fi
\usepackage[dvipsnames]{xcolor}
\usepackage{url}

\usepackage[caption=false]{subfig}
\usepackage{fontawesome}
\usepackage{lipsum}
\usepackage{comment}
\usepackage{tabularx}
\usepackage{booktabs}
\usepackage{multirow}
\usepackage[normalem]{ulem}
\usepackage{balance}
\usepackage{wrapfig}
\usepackage{todonotes}
\usepackage{calc}
\usepackage{enumitem}
\usepackage{pdfcomment}

\newlength\myheight
\newlength\mydepth
\settototalheight\myheight{Xygp}
\newcommand{\bpstart}[1]{\vspace{1mm} \noindent{\textbf{#1.}}}

\definecolor{oldDcColor}{rgb}{.988,.553,.349}
\definecolor{newDcColor}{rgb}{.60,.557,.765}
\definecolor{customBlue}{rgb}{.23,.51,.72}
\definecolor{customOrange}{rgb}{1,.5,.05}
\definecolor{customGreen}{rgb}{.17,.63,.17}
\definecolor{customPurple}{rgb}{.58,.40,.74}
\definecolor{customRed}{rgb}{.84,.15,.16}
\definecolor{customGreen}{rgb}{.17,.63,.17}

\newcommand{\edit}[1]{\textcolor{black}{#1}}
\newcommand{\added}[1]{\textcolor{black}{#1}}
\newcommand{\remove}[1]{\textcolor{brown}{\sout{}}}

\newcommand{\orgdc}[1]{\textbf{\textcolor{oldDcColor}{#1}}}
\newcommand{\newdc}[1]{\textbf{\textcolor{customBlue}{#1}}}

\ifCLASSINFOpdf
\else
\fi
\begin{document}
\bstctlcite{IEEEexample:BSTcontrol}
%
\title{Interweaving Multimodal Interaction \edit{with} Flexible Unit Visualizations for Data Exploration}
%
%
%
%

\author{Arjun Srinivasan, Bongshin Lee, and John Stasko
\IEEEcompsocitemizethanks{\IEEEcompsocthanksitem Arjun Srinivasan and John Stasko are with Georgia Institute of Technology, Atlanta,
GA, 30332.\protect\\
E-mail: {arjun010, stasko}@cc.gatech.edu
\IEEEcompsocthanksitem Bongshin Lee is with Microsoft Research, Redmond, WA, 98052.\protect\\
E-mail: bongshin@microsoft.com}
\thanks{Manuscript received April 19, 2005; revised August 26, 2015.}}

%
%

\markboth{Journal of \LaTeX\ Class Files,~Vol.~14, No.~8, August~2015}%
{Shell \MakeLowercase{\textit{et al.}}: Bare Demo of IEEEtran.cls for Computer Society Journals}
%



\IEEEtitleabstractindextext{%
\begin{abstract}
Multimodal interfaces that combine direct manipulation and natural language have shown great promise for data visualization.
Such multimodal interfaces allow people to stay in the flow of their visual exploration by leveraging the strengths of one modality to complement the weaknesses of others.
In this work, we introduce an approach that interweaves multimodal interaction combining direct manipulation and natural language with flexible unit visualizations.
We employ the proposed approach in a proof-of-concept system, DataBreeze.
Coupling pen, touch, and speech-based multimodal interaction with flexible unit visualizations, DataBreeze allows people to create and interact with both systematically bound (e.g., scatterplots, unit column charts) and manually customized views\edit{, enabling a novel visual data exploration experience.}
\remove{This helps people construct rich mental models of an information space, enabling a novel visual data exploration experience.}
We describe our design process along with DataBreeze's interface and interactions, delineating specific aspects of the design that empower the synergistic use of multiple modalities.
We also present a preliminary user study with DataBreeze, highlighting the data exploration patterns that participants employed.
Finally, reflecting on our design process and preliminary user study, we discuss future research directions.
\end{abstract}

}

\maketitle

\IEEEdisplaynontitleabstractindextext

%
\IEEEpeerreviewmaketitle

\IEEEraisesectionheading{\section{Introduction}\label{sec:introduction}}

%
%
%
%

\IEEEPARstart{R}{}ecently, there has been increased interest within the visualization community to investigate new interaction experiences, many emerging from non-traditional input devices and modalities~\cite{elmqvist2011fluid,lee2012beyond,roberts2014visualization,badam2017affordances,lee2018multimodal}. While initial efforts focused on exploring the use of individual input modalities, more recent efforts have begun to examine how multiple forms of input can be combined together to support more naturalistic interactions. Such {\em multimodal} interfaces offer great potential for data visualization, allowing people to stay in the flow of their visual exploration by leveraging the strengths of one interaction modality to complement the weaknesses of others~\cite{lee2018multimodal}. A series of recent research projects have investigated multimodal input for data visualization and shown that supporting direct manipulation (DM) and natural language (NL) input together can enhance the user experience and improve system usability~\cite{gao2015datatone,setlur2016eviza,srinivasan2018orko,kassel2018valletto}.

However, these efforts have focused on exploring NL-first interactions, using DM to overcome ambiguity in NL~\cite{gao2015datatone,setlur2016eviza} or to refine the results of NL commands~\cite{kassel2018valletto,srinivasan2018orko}.
While clearly valuable, these approaches impose a higher reliance on NL, narrowing the possible space of interactions and operations that one may perform.
On the other hand, work in the broader HCI community (e.g.,~\cite{bolt1980put,hauptmann1989speech,laput2013pixeltone,cohen1989synergistic,cohen1997quickset-2,kim2019vocal}) has shown that multimodal interactions that synergistically combine DM and NL can help design post-WIMP interfaces where the ``interface disappears," enabling people to naturally perform desired operations without solely relying on conventional graphical widgets such as menus and icons (or buttons)~\cite{van1997post}.


\edit{Our goal in this research is to explore whether that same synergy can be brought to data visualization. By combining DM and NL more deeply, can we create visualization interfaces that allow people to interact with data more fluidly and naturally? Furthermore, what types of visual representations and interactive operations facilitate such a synergistic interface? To address these research questions, we introduce an approach that interweaves \textit{DM- and NL-based multimodal interaction} with a \textit{flexible visual representation} to enable a novel visual data exploration experience.}

\edit{As an initial example, we focus on unit visualizations as the underlying visual representation.
Unit visualizations are ``visualizations that maintain the identity property of its visual marks, i.e., where each visual mark is a unique entity that is associated with a corresponding unique data item"~\cite{park2018atom}.
By representing individual data items as unique visual marks, unit visualizations support both gaining an overview of the data space as well as allowing for item-level interactions, such as querying and filtering individual data points~\cite{drucker2015unifying,park2018atom}.}
To help people construct mental models of an information space, we allow them to freely position, color, and order marks \edit{in unit visualizations}.
This freedom allows people to create not only systematically bound views (e.g., scatterplots, unit column charts), but also manually customized views based on their external knowledge of the data or subjective criteria.
Such flexibility has been shown to be valuable to analysts during sensemaking and exploratory data analysis\cite{shipman1999,andrews2010space}.

We posit that interacting with individual marks in such flexible unit visualizations is natural via DM, while operations on a group of marks (e.g., changing the properties of points that satisfy given criteria) are better performed with speech.
Correspondingly, we employ multimodal interaction that combines DM (through pen and touch) and NL (through speech) to create and interact with both systematically bound and manually customized views.

We operationalize our proposed approach in a proof-of-concept visualization system, \textit{DataBreeze}.
With each data item as a circle mark, DataBreeze initially presents the entire dataset as a cluster in a circular shape. People can then create and interact with desired views using pen, touch, and speech.
Our motivating usage scenario exemplifies how the combination of flexible unit visualizations and multimodal interaction in DataBreeze supports \remove{a novel visual data exploration experience}\edit{free-form visual data exploration}. To assess our design and understand how people employ the proposed approach, we conducted a preliminary user study where six participants used DataBreeze to explore a U.S. colleges dataset to shortlist ones of personal interest. We observed that participants inspected the data in novel ways, adopting varying data exploration patterns.
Finally, reflecting on our design process and preliminary user study, we discuss potential directions for future research.

In summary, the primary contributions of this paper are:
\begin{itemize}[leftmargin=.2in]
    \item We introduce an approach that interweaves multimodal interaction combining DM (through pen and touch) and NL (through speech) with flexible unit visualizations to facilitate a novel visual data exploration experience.
    \item We present DataBreeze, a proof-of-concept system realizing our proposed approach.
    We discuss our iterative design process as well as specific aspects of the system interface and interaction that empower synergistic use of the different modalities.
    \item We report findings from a preliminary user study with DataBreeze, highlighting the use of different input modalities and the different data exploration patterns that participants employed.
\end{itemize}
\section{Related Work}

\subsection{Pen and Touch Interaction for Data Visualization}

A large body of work has investigated the use of pen and touch interaction to design post-WIMP visualization interfaces.
While some of these efforts have explored pen-only or touch-only interaction with visualization systems (e.g.,~\cite{schmidt2010set,browne2011data,baur2012touchwave,drucker2013touchviz,rzeszotarski2014kinetica,sadana2014designing}), others have also demonstrated how bimanual interaction combining pen and touch can lead to novel or enhanced interaction experiences (e.g.,~\cite{frisch2009investigating,frisch2010diagram,lee2013sketchstory,lee2015sketchinsight,jo2015wordleplus,zgraggen2014panoramicdata,jo2017touchpivot,romat2019activeink}).

\edit{Frisch et al.~\cite{frisch2009investigating,frisch2010diagram} investigated how people use pen and touch to support both structural editing and freehand sketching to edit node-link diagrams.
In addition to eliciting gestures, they also highlight interaction design challenges to consider when combining pen and touch input.}
SketchStory~\cite{lee2013sketchstory} demonstrates how free-form sketching with a pen coupled with simple touch interactions can be leveraged to create engaging data-driven presentations.
\remove{WordlePlus~\cite{jo2015wordleplus} enables the creation of dynamic and engaging wordles by supporting direct manipulation on words with pen and touch interaction.}PanoramicData~\cite{zgraggen2014panoramicdata} and SketchInsight~\cite{lee2015sketchinsight} show how a combination of pen and touch can enable more naturalistic data exploration on an infinite canvas.
\remove{TouchPivot~\cite{jo2017touchpivot} illustrates how pen and touch input can be coupled with elements of WIMP interfaces to help novices conduct visual data exploration on tablet devices.}As a more recent example, ActiveInk~\cite{romat2019activeink} shows how pen and touch can support seamless switching between data exploration and externalization to facilitate sensemaking.
Although our work also supports pen- and touch-based input for visual data exploration, our focus is not on designing new gestures that leverage pen and touch.
Instead, we place more emphasis on exploring multimodal interactions that combine pen, touch, and speech, while taking into account the underlying visual representation of the data.

\subsection{Natural Language-Based Visualization Systems}

A range of commercial (e.g.,~\cite{IBMWatson,mspowerbi,tableauaskdata}) and research-oriented (e.g.,~\cite{cox2001multi,sun2010articulate,gao2015datatone,setlur2016eviza,aurisano2016articulate2,hoque2018applying,srinivasan2018orko,kassel2018valletto,setlur2019inferencing}) systems have investigated the use of NL input for data visualization.
For instance, Cox et al.~\cite{cox2001multi} demonstrated how explicit NL commands and dialogue can be used to specify visualizations.
In addition to NL, their system supports DM interaction (e.g., selection) with specified visualizations.
Articulate~\cite{sun2010articulate} maps user queries to tasks and uses these tasks in combination with data attributes to generate visualizations corresponding to those queries.
DataTone~\cite{gao2015datatone} illustrates how NL queries can be ambiguous when specifying visualizations and presents a mixed-initiative interface to resolve these ambiguities through GUI widgets.
Alternatively, systems like Eviza~\cite{setlur2016eviza} and Evizeon~\cite{hoque2018applying} have placed emphasis on conversational interaction, showcasing how NL can help people preserve a visual analytic flow.
\remove{Another recent example, Valletto~\cite{kassel2018valletto} demonstrates how people can use speech to create a visualization and then refine it (e.g., rotate, change mark types) using touch gestures.}\edit{Other recent multimodal systems including Orko~\cite{srinivasan2018orko} and Valletto~\cite{kassel2018valletto} also demonstrate how touch and speech can be used for visual data exploration.
While Orko~\cite{srinivasan2018orko} illustrates how multimodal interaction can aid visual exploration of network data using node-link diagrams, Valletto~\cite{kassel2018valletto} highlights how one can use speech to create a visualization and then refine it (e.g., rotate, change mark types) using touch gestures.}

\edit{As part of our work, we also support NL-based interaction, building upon current techniques for resolving ambiguity~\cite{gao2015datatone,setlur2016eviza} and supporting pragmatics~\cite{hoque2018applying,srinivasan2018orko,setlur2019inferencing}.
Advancing the line of research on multimodal visualization systems supporting NL input, we examine the use of an additional input modality (pen) with a more general category of visual representations (unit visualizations).
In doing so, we explore additional types of multimodal interactions for a wide range of general interactive visual analysis tasks~\cite{heer2012interactive}
including view specification (e.g., changing axes, ordering), view manipulation (e.g., explicitly coloring or moving points), and externalization of one's exploration process (e.g., inking, labeling).}
\remove{Most related to our work is Orko~\cite{srinivasan2018orko} that supports touch- and speech-based multimodal input for visual exploration of network data.
As an initial example of a multimodal visualization system using speech, Orko facilitates the high-level task of interacting with a node-link diagram, supporting a small set of operations for common network visualization tasks (e.g., finding paths between two nodes).
We advance this line of research on multimodal visualization systems by examining the use of an additional input modality (pen) with unit visualizations.
In doing so, we explore additional types of multimodal interactions for a wide range of general interactive visual analysis tasks~\cite{heer2012interactive} that go beyond Orko, involving view specification (e.g., changing axes, ordering), view manipulation (e.g., explicitly coloring or moving points), and externalization of one's exploration process (e.g., inking, labeling).
Our work also substantially extends Orko's NL capabilities by 1) enabling additional types of sequential and simultaneous multimodal commands, 2) supporting follow-up commands for a wider range of operations, and 3) inferring potential reasons for system errors and providing example commands to help users correct those errors.}

\subsection{Unit Visualizations with Naturalistic Interactions}

\edit{We chose unit visualizations as an initial example because they have frequently been used as 
visual representations for data exploration in systems that support more naturalistic forms of input (e.g.,~\cite{heilig2010scattertouch,rzeszotarski2014kinetica,sadana2014designing,dai2015hands,xia2018dataink,le2019dynamic}).}
For example, Rzeszotarski and Kittur presented Kinetica~\cite{rzeszotarski2014kinetica}, a tablet-based visualization tool for exploring multivariate data.
They described how multi-touch gestures coupled with physics-based affordances can let people fluidly perform operations such as specifying axes and filtering.
They also discussed how such naturalistic interactions with unit visualizations help people build rich mental models of an information space by keeping data salient and enabling tracking of data points during exploration.
\edit{With ScatterTouch~\cite{heilig2010scattertouch}, Heilig et al. illustrated how simple multi-touch gestures can be leveraged to interactively create focus regions in a scatterplot, supporting co-located data exploration.}
Sadana and Stasko~\cite{sadana2014designing} presented a tablet-based visualization system, illustrating how multi-touch gestures can be used to perform various operations with scatterplots.
Dai et al.~\cite{dai2015hands} implemented a version of the Dust \& Magnet technique~\cite{yi2005dust} on a large touch display, leveraging its multi-touch capabilities.
It extended the original technique by allowing the simultaneous manipulation of multiple magnets, enabling more expressive and fine-tuned interaction.

While we build upon the common notion of unit visualizations, we investigate a variant that affords more flexibility in its specification and manipulation.
\edit{We illustrate how coupling this variant with multimodal interaction enables a novel free-form visual data exploration experience.}
\section{Synergistic Coupling of Flexible Unit Visualizations and Multimodal Interaction}

\begin{figure}[b!]
    \centering
    \includegraphics[width=\linewidth]{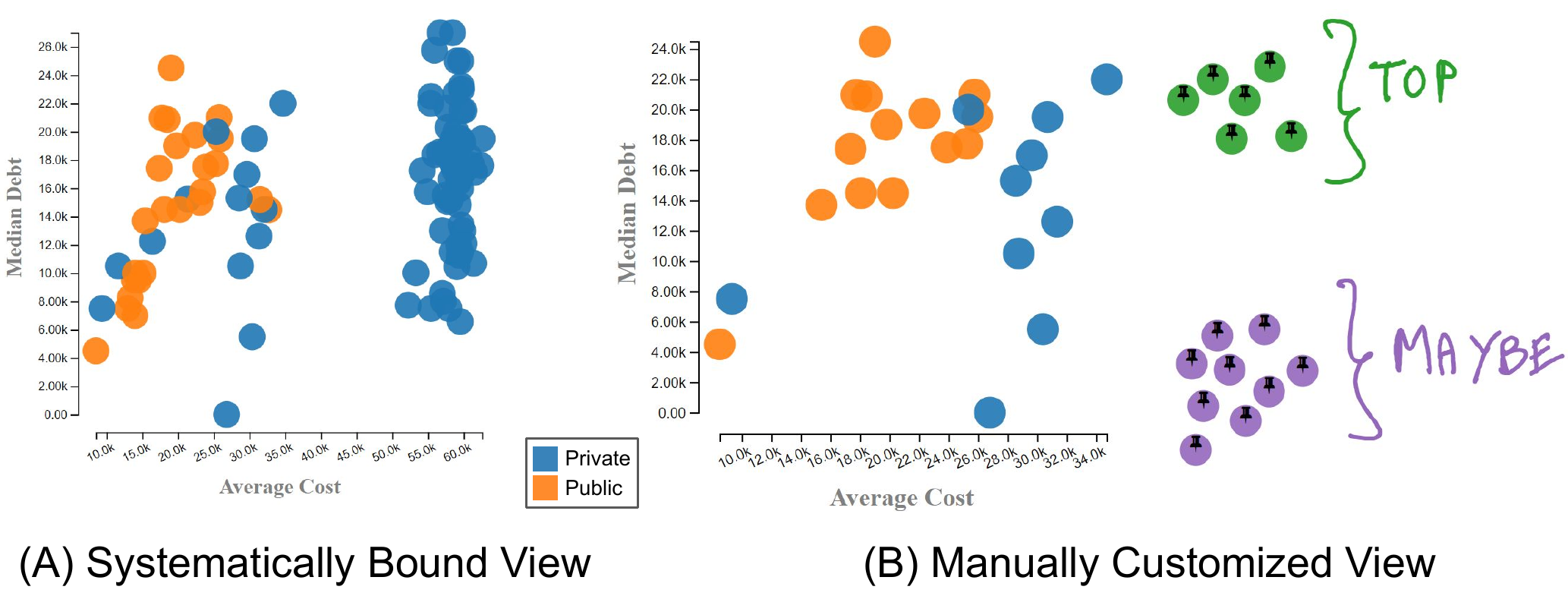}
    \vspace{-3mm}
    \caption{Flexibility afforded by unit visualizations.
    Both visualizations are displaying a U.S.~colleges dataset.
    (A) A standard systematically bound scatterplot showing the relationship between \textit{Average Cost} and \textit{Median Debt}.
    (B) A manually customized view having a scatterplot (which is still bound to \textit{Average Cost} and \textit{Median Debt}) and two groups of preferred and potential colleges \added{created by explicitly moving and coloring points from the initial scatterplot.}
    }
    \label{fig:flexibility-spectrum}
\end{figure}

By representing individual data cases as unique visual marks, unit visualizations allow people to specify views with different levels of customization.
For instance, a 2D scatterplot is an example of a
\textit{systematically bound} view, where the position of each visual mark is strictly bound to two data attributes (e.g., Figure~\ref{fig:flexibility-spectrum}A).
Alternatively, by explicitly changing the properties (e.g., position, color) of marks, one can create a \textit{manually customized} view, where not all points in the view are bound to the same set of data attributes (e.g., Figure~\ref{fig:flexibility-spectrum}B).
To allow users to create both systematically bound and manually customized views, we propose the notion of  \textbf{flexible unit visualizations}.

Similar to flexible linked axes~\cite{claessen2011flexible}, flexible unit visualizations could be especially powerful during more open-ended data exploration.
For example, users could apply well-known unit visualization layouts (e.g., scatterplots, unit column charts) to correlate and compare values.
Once they identify a set of points of their interest or have a specific exploration criteria in mind, users could then explicitly move points out from the scatterplot into separate groups, coloring these groups to create a customized view that best fits their mental model (e.g., Figure~\ref{fig:flexibility-spectrum}B).
Furthermore, the ability to spatially organize data and create virtual workspaces can also aid externalization during the sensemaking process, especially in the initial, exploratory phases~\cite{andrews2010space,andrews2013impact}.

Such flexible data exploration is difficult with existing unit visualization tools that support interaction only through DM and control panels~\cite{endert2012semantic,rzeszotarski2014kinetica,sadana2014designing,drucker2015unifying,saket2017visualization,major2019graphicle}.
For example, SandDance~\cite{drucker2015unifying} only allows systematic specification: the changes to the view need to be specified through menus, not allowing users to manipulate individual data points to adjust the view.
Alternatively, while other systems such as ForceSPIRE~\cite{endert2012semantic} and VisExemplar~\cite{saket2017visualization} support direct interaction with individual data points, changes resulting from this interaction are propagated to all data points, ultimately resulting in a systematically bound view.
For example, juxtaposing two points in ForceSPIRE recomputes the force-directed layout and repositioning points in VisExemplar creates a new scatterplot.



In general, relying solely on DM and control panels for interaction with a unit visualization can be problematic: the user interface can be overly complex when incorporating all of the desired actions and user interaction can be tedious for creating manually customized views (e.g., moving points that are not spatially clustered).
On the other hand, the complementary strengths of DM and NL make a combination of the two a potentially effective solution for this challenge.
The synergistic use of DM and NL when interacting with flexible unit visualizations could allow users to create both systematically bound and manually customized views, also enabling seamless switching between the views.
Specifically, the precision and control afforded by DM can allow people to make fine-tuned, customized changes, whereas the ability of NL to augment systematic actions can help them overcome the repetition that accompanies DM.

\subsection{Motivating Scenario: Identifying Preferences among US Colleges}

To illustrate how multimodal interaction with flexible unit visualizations can facilitate visual data exploration, we describe a usage scenario.
Imagine Sarah, a parent who is identifying colleges her daughter might want to apply to.
Sarah downloads a dataset of the top 100 schools\protect\footnote{We describe the scenario with a small set of 100 points only to improve the readability of figures. Our approach scales to larger data sets as illustrated in the supplementary video.} in the U.S.~from a popular college ranking website. 
The dataset contains 14 attributes for each college including both categorical (e.g., \textit{Region}, \textit{Locale}) and quantitative (e.g., \textit{SAT Average}, \textit{Average Cost}) attributes.
For consistency, we use this dataset in our examples throughout the paper.

\begin{figure*}[t!]
    \centering
    \includegraphics[width=\linewidth]{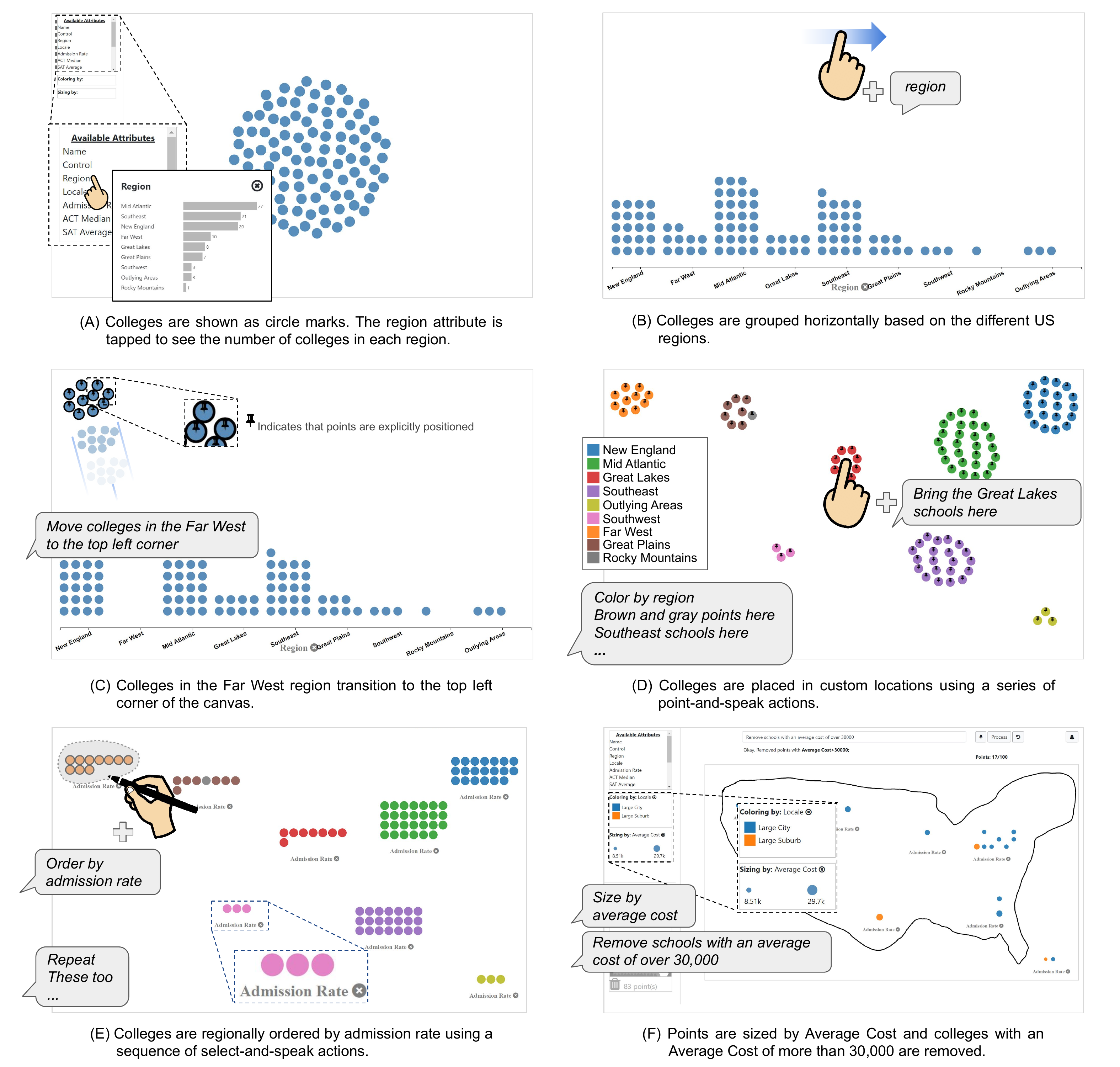}
    \caption{\edit{Scenes illustrating the usage scenario of exploring colleges in the U.S. Sub-figure captions summarize the system states}.}
    \vspace{-2mm}
    \label{fig:scenario-1}
\end{figure*}

\bpstart{Getting an overview from systematically bound views}
The system initially shows all points clustered at the center of the screen.
To learn more about the available attributes, Sarah taps on the different attributes in the attribute summary panel (Figure~\ref{fig:scenario-1}A).
Looking at the \textit{Region} attribute, Sarah decides to categorize colleges by their regions.
To do this, she swipes from left-to-right \raisebox{-.5em}{\includegraphics[height=1.5em]{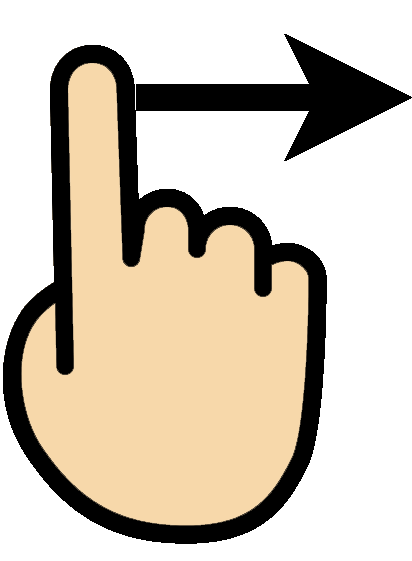}} on the canvas and says ``\textit{Region}."
Inferring the specification of an axis through the swipe gesture and identifying the attribute via speech, the system creates a column chart grouping colleges by their regions (Figure~\ref{fig:scenario-1}B).
Although this chart gives her a good overview, Sarah finds it difficult to compare the different regions because the placement of regions does not visually correspond to their geographic locations.

\bpstart{Contextualizing the data space with customized views}
Sarah decides to adjust the view so the points are positioned similar to how they look on a map of the U.S.
To do this, she starts by saying ``\textit{Move the colleges in the Far West to the top left corner.}"
This moves colleges in the \textit{Far West} region to the top left corner of the canvas (Figure~\ref{fig:scenario-1}C).
The {\small{\faThumbTack}} icon indicates that Sarah explicitly moved the colleges and their positions are no longer bound to the X-axis. 
To keep track of the different regions, Sarah says ``\textit{Color by region}"---applying a global color mapping to all points.
She then continues to reposition other points through a series of speech-only (e.g., ``\textit{Move the New England schools to the top right corner}") and multimodal utterances (e.g., \raisebox{-.5em}{\includegraphics[height=1.5em]{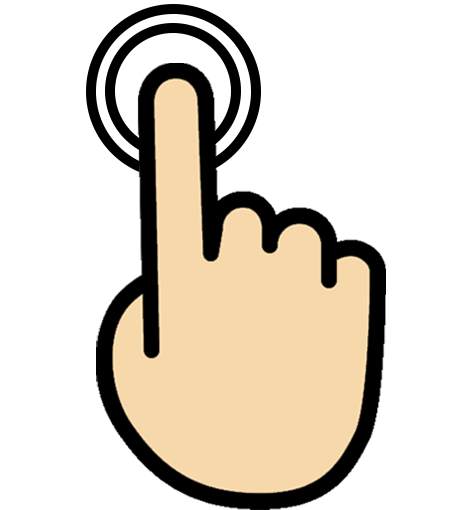}} + ``\textit{Bring the Great Lakes schools here},'' \raisebox{-.5em}{\includegraphics[height=1.5em]{figures/icons/point-finger.png}} + ``\textit{Brown and gray points here}'').
This results in a view that combines custom positioning of the colleges with the systematic color mapping based on the \textit{Region} attribute (Figure~\ref{fig:scenario-1}D).

\bpstart{Adding localized mappings for detailed exploration}
Curious about the competitiveness of schools in the different regions, Sarah decides to organize the schools further by their \textit{Admission Rate}s.
She starts from the Far West schools: after selecting the schools by drawing a lasso \raisebox{-.5em}{\includegraphics[height=1.5em]{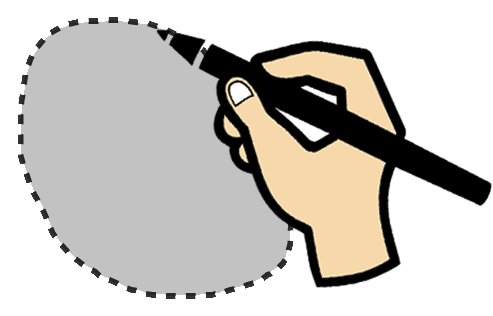}} around them, Sarah says ``\textit{Order by admission rate.}"
The system, in response, re-orders the selected Far West schools by their admission rates (in an ascending order).
Because the position of the points is now determined by an attribute value, the system removes the {\small{\faThumbTack}} icon from the re-ordered points.
Sarah repeats this select-and-order sequence for other regions by selecting groups of points and saying commands like ``\textit{repeat}" and ``\textit{these too.}"
Inferring from her previous commands, the system re-orders the colleges by their admission rates, leading to the view shown in Figure~\ref{fig:scenario-1}E.

\begin{figure*}[t!]
    \centering
    \includegraphics[width=.7\linewidth]{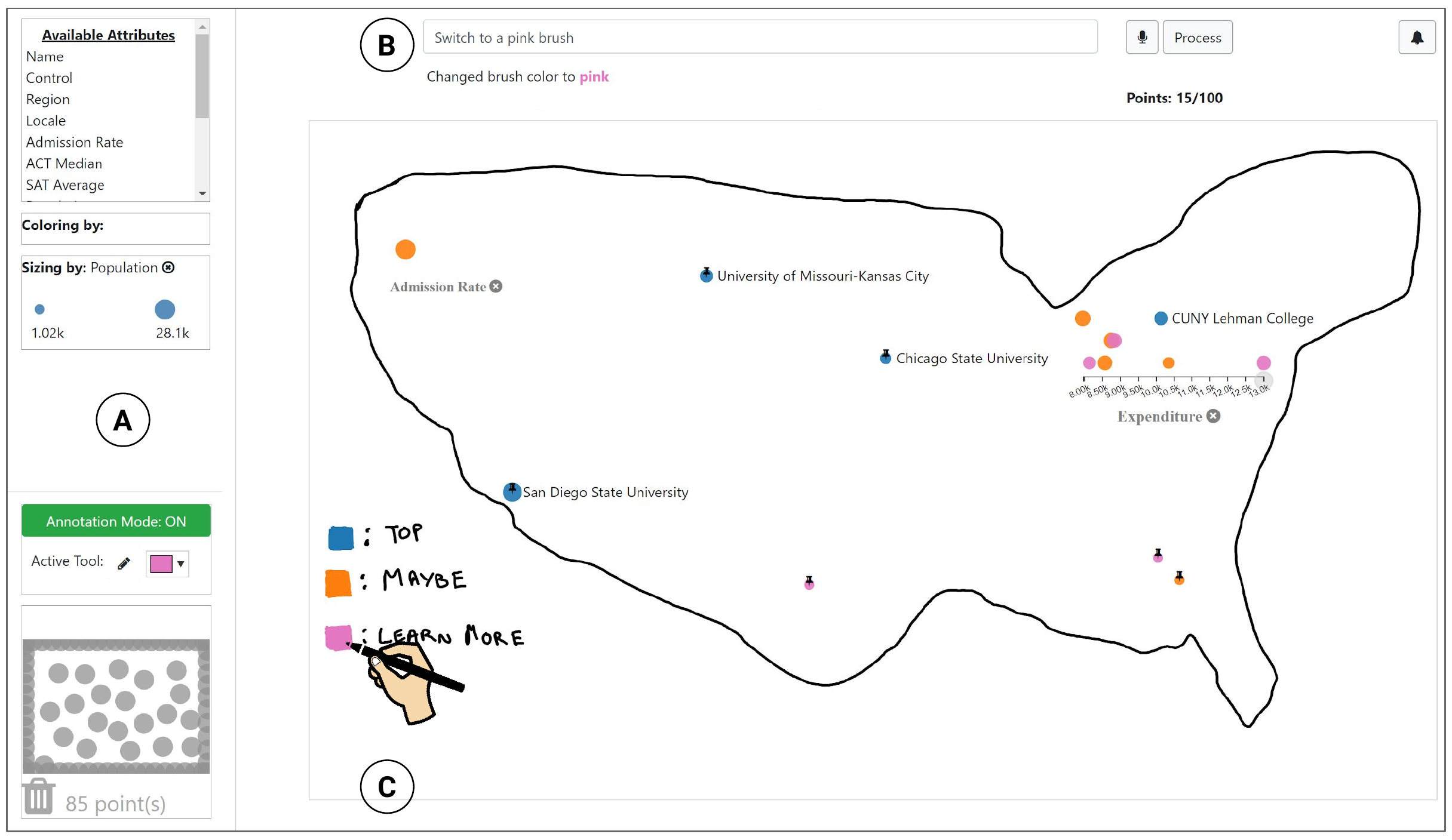}
    \caption{\edit{The system's interface as Sarah concludes her exploration: (A) Side panel with (from top to bottom) the attribute summary container, legends for global coloring and sizing attributes, annotation options, and a virtual bin for filtered points; (B) Speech input and feedback row; (C) Main canvas.
    Here, Sarah is using the brush tool to draw a custom color legend.}}
    \label{fig:scenario-end}
\end{figure*}


\bpstart{Adding annotations and global mappings to switch back to overview-level exploration}
To externalize her mental mapping of the data space on the canvas, Sarah draws a rough outline around the points using the brush tool {\small{\faPaintBrush}} to make the view look like a map of the U.S.
As they might provide better career opportunities in the future, Sarah is more interested in schools in larger localities.
To see the types of locations the schools are in, she says ``\textit{Color by locale}" and scans the updated chart.
Noticing a lot of schools are in remote locations or small towns, Sarah filters the view by saying ``\textit{Remove schools that are not in large cities or large suburbs.}"
This removes 47 points from the canvas.
Next, to get a sense of the cost to attend each college, Sarah issues the command ``\textit{Size by average cost}" and notices that except for some colleges in the Great Lakes and Rocky Mountains, most colleges are expensive and potentially beyond their budget.
To filter colleges further, she says ``\textit{Remove schools with an average cost of over 30,000}," leaving only 17 colleges on the canvas (Figure~\ref{fig:scenario-1}F).

\bpstart{Combining global and local operations for drill-down}
Sarah taps on points individually to see their details and explores the colleges from different perspectives by iteratively assigning different attributes
to the position, size, and color of points.
To enable better comparison of schools within a geographic region, Sarah also creates local views by mapping attributes to the positions of subsets of points (Figure~\ref{fig:scenario-end}), inspecting the individual groups.
Based on this inspection, Sarah decides that she does not want to consider the two colleges in the `Outlying Areas' and removes them.

\bpstart{Externalizing custom mappings}
Inspecting the remaining 15 schools, Sarah identifies four schools that are her top picks, six schools that she would strongly consider, and five schools that she needs to read more about.
To externalize this mental ranking, Sarah selects the four schools that are her top choices and says ``\textit{Add labels and color them blue}" to show their names and change their color.
Sarah then selects the six colleges that she would strongly consider and says ``\textit{Color these orange}," and selects the remaining five schools and says ``\textit{Pink.}"
Sarah again activates the brush tool and draws a legend to note her color choices (Figure~\ref{fig:scenario-end}).
Concluding her exploration, Sarah sends this view via an email to herself to discuss it with her family members.

\subsection{Design Process and Goals}
To create a system that supports the illustrated style of visual data exploration, we followed an iterative design process that helped us identify and refine a set of design goals we needed to accomplish.
Below, we describe this design process and resulting design goals.

We first define a few key terms we use throughout this paper.
We use the term \textit{command} to refer to any type of NL utterance such as a query, comment, or question.
By an \textit{operation}, we refer to actions like selection, changing encodings, coloring points, etc.
Operations typically require \textit{parameters} (e.g., attributes, color names, data values) and operate on one or more \textit{targets} (e.g., all points on the canvas, selected points, points meeting a specific data criteria).


\subsubsection{Design Process}
\begin{figure}[t!]
    \centering
    \includegraphics[width=\linewidth]{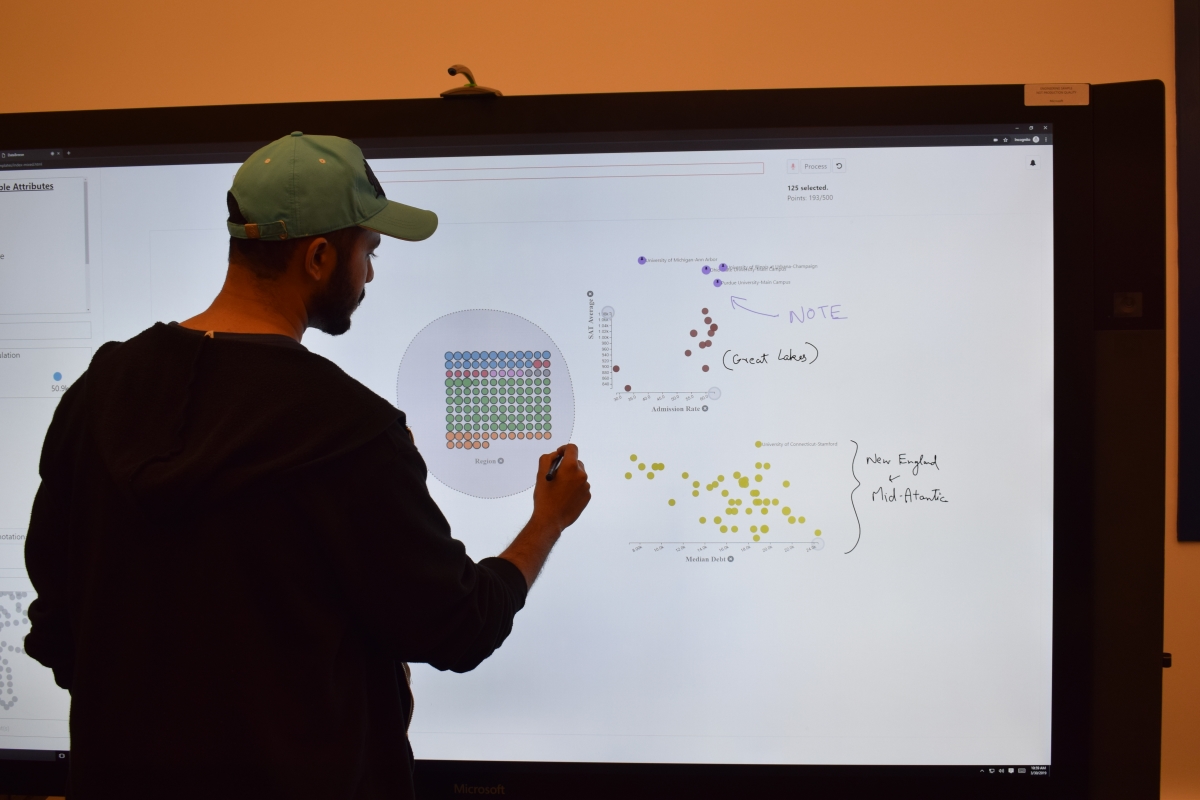}
    \caption{\added{DataBreeze running on an 84" Microsoft Surface Hub with an external microphone placed on top of the display to record speech input.}}
    \label{fig:teaser}
\end{figure}

As a test bed, we initially implemented a basic version of a pen-, touch-, and speech-based multimodal unit visualization tool.
The tool supported a minimal set of operations (e.g., assign X/Y axes, change color and size, filter) and interactions (e.g., dragging to move points, drawing a lasso for selection, speech commands for individual operations).
We implemented the system on an 84" Microsoft Surface Hub (Figure~\ref{fig:teaser}) to support scalability (in terms of number of points along with the ability to interact with individual points), as well as to provide the freedom to spatially organize the view.

We iterated on the tool's design and implementation across six design sessions (each between 30-90 minutes) among ourselves and other graduate students.
In these sessions, we investigated interactions during open-ended tasks (e.g., shortlist a set of startups to invest in, shortlist a set of colleges for your child or younger sibling). 
These design sessions allowed us to critically reflect on the design, get early feedback on the interactions (e.g., pen/touch gestures, grammar of spoken commands), and identify operations that we needed to support.
Specifically for the pen/touch gestures, due to a lack of consensus during the design sessions, we additionally conducted four informal elicitation sessions with graduate students, observing how they performed actions such as selecting and moving points, invoking and interacting with context menus, etc.

\subsubsection{Design Goals}
We considered several factors for developing a multimodal system supporting flexible unit visualizations (e.g., How should the system integrate input from multiple modalities? Should changes explicitly made to a subset of points be propagated to all points in the view?).
Below we list the goals that we initially had in mind at the start of the project (\orgdc{DG1-4}) as well as the ones we incrementally derived based on the observations during the design sessions (\newdc{DG5-8}).
While these goals are primarily applicable to multimodal interfaces supporting flexible unit visualizations, some of them (\orgdc{DG2-4}, \newdc{DG6}, \newdc{DG7}) are also generally applicable to DM- and NL-based multimodal visualization systems.


\vspace{.5em}
\noindent\textbf{\orgdc{DG1.}~Support both systematic binding and manual customization.}
The premise of this work is that allowing people to manually customize systematically bound visualizations can aid data exploration, offering high flexibility.
To enable this manual customization, the system should allow users to specify visualizations similar to current tools (e.g., assigning X/Y-axes attributes to create a scatterplot), while still supporting data item- or subset-level manipulation (e.g., dragging points out of a scatterplot to create a customized group, changing the color of specific points).

\vspace{.5em}
\noindent\textbf{\orgdc{DG2.}~Support various multimodal input patterns.}
Prior research on multimodal interfaces has shown that although a system supports multiple modalities, people may choose to interact using a single modality and not combine inputs~\cite{oviatt1999ten}.
Furthermore, even when using multiple modalities, people may not use them simultaneously and instead combine them sequentially (e.g., select a set of points with touch, pause, and then issue a spoken command)~\cite{oviatt1999ten,oviatt1997integration}.
Hence, the system should support unimodal input as well as both sequential and simultaneous integration of modalities.

\vspace{.5em}
\noindent\textbf{\orgdc{DG3.}~\remove{Avoid complex}\edit{Leverage simple} pen/touch gestures.}
While pen/touch input can be highly expressive, complex gestures involving multiple fingers or bimanual interaction can be difficult to learn and discover~\cite{wu2006gesture}.
These challenges are further amplified with the addition of a third modality in the form of speech.
Therefore, the system should leverage simple and familiar pen/touch interactions that are easy to learn while still supporting the required set of operations.

\vspace{.5em}
\noindent\textbf{\orgdc{DG4.}~Provide instruction and feedback for speech input.}
Lack of instruction (knowing what can be said) and feedback (understanding what the system did in response to a command) are well-known challenges with NL interaction~\cite{brennan1995interaction,yankelovich1995designing,yankelovich1996users,shneiderman1997direct}.
This challenge is amplified in multimodal interfaces where the linguistic structure of commands may differ when users interact multimodally~\cite{oviatt1999ten}.
Hence, the system should assist discovery and learning of the supported range of speech commands.
Furthermore, the system should make users aware of the actions it took in response to speech commands---giving users the option to revert or correct them.

\vspace{.5em}
\noindent\textbf{\newdc{DG5.}~Support both global and local changes.}
We observed that participants wanted to perform operations at both a global (e.g., creating a scatterplot using all data points) and a local level (e.g., selecting a subset of points within a scatterplot to form an ordered group).
Therefore, to support incremental data exploration and smoother transitions between systematically bound and customized views (\orgdc{DG1}), the system should let users perform operations on all points on the canvas (global operations) or on a subset of data items (local operations).
However, local changes may conflict with previously applied global mappings (e.g., moving points may break a globally specified position mapping).
To overcome this, we initially removed the global mapping in case of conflicting local changes (e.g., removing the global X-axis scale if a set of points are explicitly moved).
During the design sessions, however, we observed that participants preferred that visual elements of the previously applied global changes be preserved even when conflicting local changes are made.
We found that this helped participants contextualize changes and continue their exploration.
Hence, the system should try to preserve context for local changes and provide visual cues to differentiate between local and global mappings.
For instance, in Figure~\ref{fig:scenario-1}C, the \textit{Region} scale is shown on the global X-axis even though a subset of the points are moved out.
Furthermore, the {\small{\faThumbTack}} icons on the moved points indicate that the points are not bound to the global view.

\vspace{.5em}
\noindent\textbf{\newdc{DG6.}~Support equivalence between pen and touch.}
Following Hinckley et al.'s guideline of ``\textit{pen writes, touch manipulates,}"~\cite{hinckley2010pen+} we initially applied a division of labor tactic separating the roles of pen and touch: we let people draw lassos and select points using the pen, while moving the points with a finger.
During the design sessions, however, we observed that participants frequently confused the role of pen and touch, often trying to use the two interchangeably (e.g., drag points with a pen, select points with a finger).
This adversely affected the system's usability, suggesting that the benefits of greater expressiveness through separated roles was not worth the resulting confusion it caused.
Thus, when semantically meaningful differences are missing between pen and touch interaction, the system should support equivalent and consistent operations between the two.

\vspace{.5em}
\noindent\textbf{\newdc{DG7.}~Support implicit and explicit triggering of speech.}
While we supported unimodal, sequential, and simultaneous integration of modalities in our initial prototypes (\orgdc{DG2}), similar to prior systems~\cite{srinivasan2018orko,kassel2018valletto}, users had to explicitly trigger speech input using a ``listen'' button or a wake-word.
However, we observed that this impeded multimodal interaction often resulting in participants saying a command after performing a gesture only to realize that the system was not listening.
Therefore, to facilitate more seamless multimodal interaction, the system should provide both explicit and implicit speech activation techniques.

\vspace{.5em}
\noindent\textbf{\newdc{DG8.}~Support externalization of custom mappings.}
During the design sessions, participants created custom mappings (\orgdc{DG1}) fitting their mental models by making local changes to the view (\newdc{DG5}).
To let people externalize their custom mappings (e.g., adding labels for ``virtual bins" or drawing custom legends as in Figures~\ref{fig:flexibility-spectrum}B and ~\ref{fig:scenario-end}), the system should support basic inking features.

\section{DataBreeze}

With the design goals listed above in mind, we developed DataBreeze---a multimodal system that facilitates visual data exploration by supporting constructing and interacting with flexible unit visualizations.
Table~\ref{tbl:operations} illustrates operations currently supported in DataBreeze that were derived and refined based on the aforementioned design sessions.

\begin{table*}[t!]
\caption{Examples of supported operations and their corresponding speech and multimodal commands in DataBreeze}
\centering
\resizebox{\textwidth}{!}{%
\begin{tabular}{@{}ll@{}}
\toprule
\multicolumn{1}{c}{\textbf{Operations}} & \multicolumn{1}{c}{\textbf{Sample Speech \& Multimodal Commands}} \\ 
\midrule
\textbf{Assign X/Y-axes} & \textit{Sort vertically by Admission Rate; \raisebox{-\mydepth}{\includegraphics[height=1.5em]{figures/icons/swipe-lr-finger.png}} + SAT Average; Align horizontally by debt;} \\
\textbf{Filter} & \textit{Remove schools in the Far West; Remove all points except the blue ones; \raisebox{-\mydepth}{\includegraphics[height=1.5em]{figures/icons/lasso-pen-custom.png}} + Remove;} \\
\textbf{Color/Size by attribute} & \textit{Color by region; \raisebox{-\mydepth}{\includegraphics[height=1.5em]{figures/icons/lasso-pen-custom.png}} + Size these by expenditure; Color by locale and then size by average cost;} \\
\textbf{Order by attribute} & \textit{\raisebox{-\mydepth}{\includegraphics[height=1.5em]{figures/icons/lasso-pen-custom.png}} + Order by admission rate; Rearrange schools in the Southeast by their population;} \\
\textbf{Move} & \textit{Put the public schools on the right;} \textit{\raisebox{-\mydepth}{\includegraphics[height=1.5em]{figures/icons/point-finger.png}} + Bring the private schools here; \raisebox{-\mydepth}{\includegraphics[height=1.5em]{figures/icons/point-finger.png}} + Green here;} \\
\textbf{Others} & \textit{\raisebox{-\mydepth}{\includegraphics[height=1.5em]{figures/icons/lasso-pen-custom.png}} + Color red; Highlight Stanford; \raisebox{-\mydepth}{\includegraphics[height=1.5em]{figures/icons/lasso-pen-custom.png}} + Summarize; Add labels to all public schools;} \\ 
\bottomrule
\end{tabular}%
}
\label{tbl:operations}
\end{table*}

\subsection{Pen \& Touch Interaction}
\begin{figure}[t!]
    \centering
    \includegraphics[width=\linewidth]{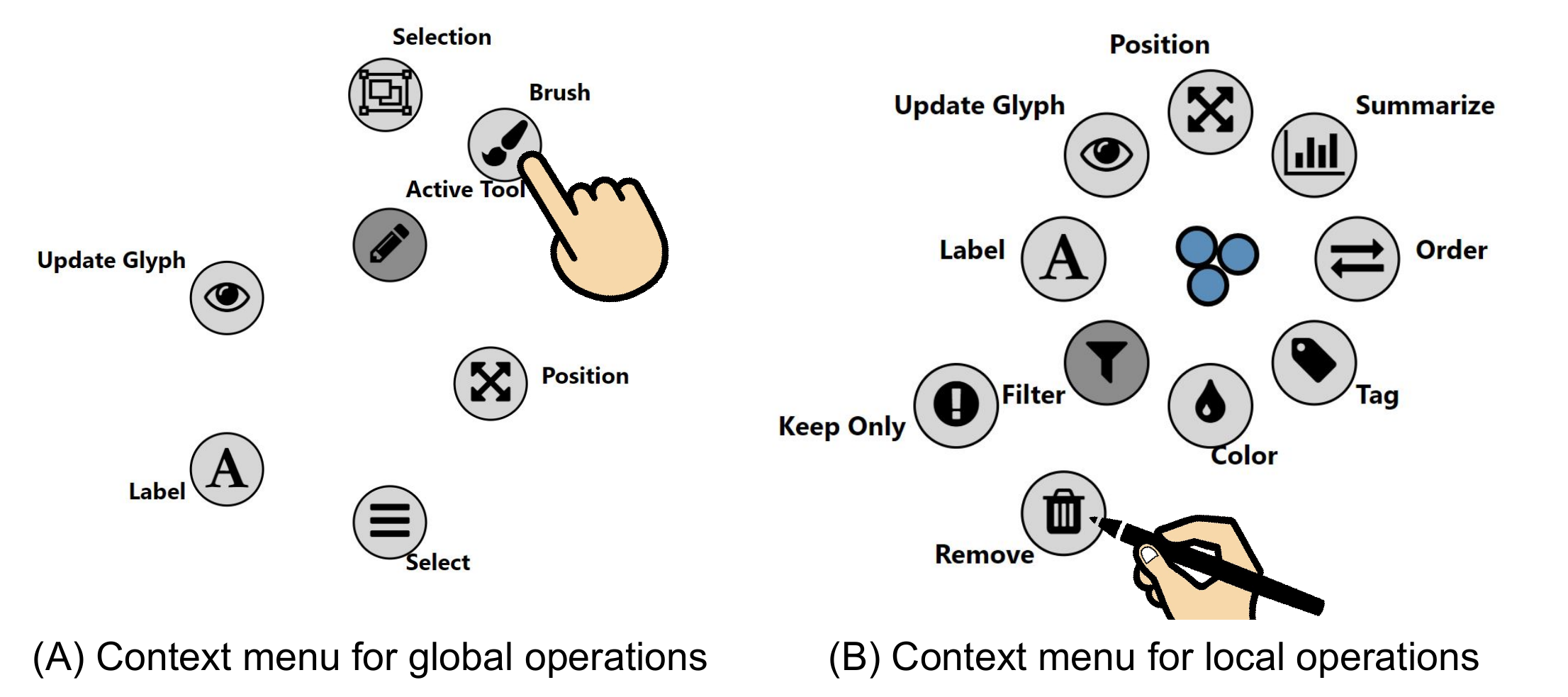}
    \caption{Context menus
    \edit{for pen/touch interaction.
    (A) A global menu is invoked by long pressing on the canvas background. The user is switching to the brush tool.
    (B) A local menu is invoked by long pressing on selected data points. The user is filtering out the selected points.
    }
    }
    \label{fig:context-menus}
\end{figure}

DataBreeze supports three familiar gestures (tap, long press, and drag) (\orgdc{DG3}) that can be performed on the canvas or on a data point (Table~\ref{tbl:pen-and-touch}).
To support externalization of custom mappings (\newdc{DG8}), DataBreeze also supports drawing
using a brush tool.
If the brush tool is active, dragging the pen on the canvas renders ink strokes.
Otherwise, pen and touch can be used interchangeably (\newdc{DG6}).
Akin to previous pen- and touch-based visualization tools~\cite{xia2018dataink,kim2019datatoon}, DataBreeze employs radial context menus (Figure~\ref{fig:context-menus}) with which people can perform operations.


\subsection{Speech Interaction}
DataBreeze allows the use of speech unimodally or as part of multimodal interactions to perform the supported operations (\orgdc{DG2}).
Table~\ref{tbl:operations} highlights some examples of supported NL-only and multimodal NL commands.

\subsubsection{Triggering Speech Input (\newdc{DG7})}

DataBreeze starts listening or recording user utterances in response to one of five user actions: 1) tapping the microphone icon ({\small{\faMicrophone}}), 2) double-tapping on the canvas (similar to knocking on a door), 3) long pressing on the canvas or a data point with a finger or pen, 4) selecting one or more points by drawing a lasso, or 5) swiping horizontally or vertically on the canvas to specify an axis.
The first two triggering techniques are explicit and allow users to initiate speech input on-demand.
The latter three are more implicit triggering techniques to support smoother multimodal interaction where the system starts listening based on the user's pen or touch actions.
Whenever the system is listening, the microphone icon and the input box flash red ({\textcolor{red}{\small{\faMicrophone}}}).


\subsubsection{Interpreting Speech Commands}

\begin{table}[t!]
\caption{Pen and touch interactions in DataBreeze. Except when the brush tool is active, pen and touch can be used interchangeably.}
\centering
\resizebox{\linewidth}{!}{%
\begin{tabular}{@{}llll@{}}
\toprule
\textbf{Gesture} & \textbf{Target} & \multicolumn{1}{c|}{\textbf{Touch}} & \multicolumn{1}{c}{\textbf{Pen}} \\ \midrule
\multirow{2}{*}{Tap} & Canvas & \multicolumn{2}{l}{Clears selections} \\
 & Point & \multicolumn{2}{l}{Shows tooltip with label} \\ \midrule
\multirow{2}{*}{\begin{tabular}[c]{@{}l@{}}Long press\\ (\textgreater{}1 sec.)\end{tabular}} & Canvas & \multicolumn{2}{l}{Select + Context menu for global operations} \\
 & Point & \multicolumn{2}{l}{Select + Context menu for local operations} \\ \midrule
\multirow{3}{*}{Drag} & Canvas & \multicolumn{1}{l|}{\begin{tabular}[c]{@{}l@{}}Draws a selection lasso\\ or initiates X/Y axis\end{tabular}} & \begin{tabular}[c]{@{}l@{}}(w/ brush tool)\\ Draws ink strokes\end{tabular} \\
 & Point & \multicolumn{2}{l}{Moves point(s)} \\
\bottomrule
\end{tabular}%
}
\label{tbl:pen-and-touch}
\end{table}

\textbf{General command interpretation strategy.}
We use a combination of a template- and lexicon-based parser to interpret speech.
DataBreeze identifies the operations, targets, and parameters of the spoken command by comparing the input to predefined command phrasing templates \edit{(e.g., \textit{Size by [attribute]})}\added{~compiled from the initial design sessions}.
If the input does not match a template, the system tokenizes the command string and compares the tokens to the system lexicon
to infer the operations, targets, and parameters.
\edit{The system lexicon contains keywords/phrases mapping to the different operations (e.g., `order,' `color,' `remove') and parameter values (e.g., attribute names and values, color names, canvas regions like `top' and `right').}

Consider the example command ``\textit{Remove all private schools with an average cost of more than 30,000}" that does not match a predefined phrasing template.
To interpret this command, the system first removes all stopwords from the input command.
Next, the system derives all N-grams from the string and compares the N-grams to the available lexicon using the cosine similarity and the Wu-Palmer similarity score~\cite{wu1994verbs}.
This comparison results in the system matching the N-gram ``\textit{private schools}'' to the attribute-value pair \textit{Control$=$Private} and ``\textit{average cost more than 30,000}'' to the attribute-value pair \textit{Average Cost$>$30,000}.
Furthermore, using the word ``\textit{Remove}'', the system also infers that the user is referring to the \textit{filter} operation.
Combining the identified operation and attribute-value pairs, the system removes all points having \textit{Control$=$Private} and \textit{Average Cost$>$30,000}.

\begin{figure}[t!]
    \centering
    \includegraphics[width=\linewidth]{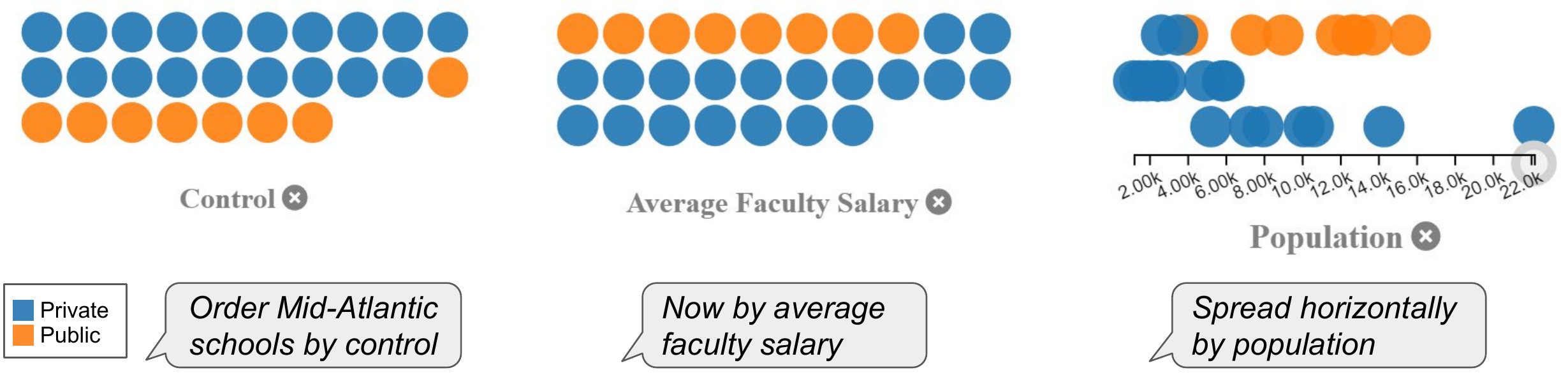}
    \caption{An interaction sequence illustrating follow-up commands in DataBreeze. The first command orders \textit{Mid-Atlantic} schools by the \textit{Control} type \added{(\textit{Public} vs.~\textit{Private})}. The subsequent command implicitly refers to the \textit{Mid-Atlantic} schools and the \textit{Order} operation from the initial command. The second follow-up command again implicitly refers to \textit{Mid-Atlantic} schools but specifies a different operation (\textit{Assigning X-axis}).}
    \label{fig:follow-up}
\end{figure}

\vspace{.5em}
\noindent\textbf{Handling follow-up commands.}
In addition to fully-specified speech commands, DataBreeze also supports follow-up commands.
\edit{We use conversational centering~\cite{grosz1995centering} to infer follow-up commands, extending the model beyond attribute-focused operations (e.g., filtering and changing encodings~\cite{hoque2018applying,srinivasan2018orko}) to support additional data item-level operations introduced in DataBreeze (e.g., updating point colors, moving points, specifying local axes).}
At a high-level, when it executes a command, the system creates a context object that contains references to the operations, parameters, and targets associated with the command.
If the subsequent command contains new operations, parameters, or targets, this context object is refreshed.
If not, the system tries to identify the missing information using the context object and updates it accordingly.
Figure~\ref{fig:follow-up} shows a sample follow-up command sequence employing this strategy.

\subsection{Multimodal Interaction}

DataBreeze supports three types of multimodal commands where operations, parameters, and targets are derived using a combination of the pen/touch and speech input.

\bpstart{Point-and-speak} 
Figure~\ref{fig:scenario-1}D shows an example of a multimodal point-and-speak command where the user points \raisebox{-.5em}{\includegraphics[height=1.5em]{figures/icons/point-finger.png}} on the canvas and says ``\textit{Bring the Great Lakes schools here}."
In this case, the command contains references to the operation (\textit{Move} operation identified using the word ``\textit{Bring}") and target (points with \textit{Region$=$Great Lakes}).
However, the parameter for the move operation (i.e., canvas location to move points to) is provided via touch and is deictically referenced using the word ``\textit{here}.''
Thus, by considering both the input command and how it was triggered, the system moves the target points to the requested location.

\bpstart{Select-and-speak}
With select-and-speak commands, users can select a set of points and issue a speech command to perform a local operation on those points.
An example of this is shown in Figure~\ref{fig:scenario-1}E where the user selects a set of points \raisebox{-.5em}{\includegraphics[height=1.5em]{figures/icons/lasso-pen-custom.png}} and says ``\textit{Order by admission rate}."
In this case, the system identifies the operation (\textit{Order}) and parameter (\textit{Admission Rate}) using the command, inferring the target based on the preceding selection that triggered speech input.

\bpstart{Swipe-and-speak}
With swipe-and-speak commands, users can swipe horizontally or vertically on the canvas and say an attribute name to position points by a specific attribute.
An example of this is shown in Figure~\ref{fig:scenario-1}B where the user swipes from left-to-right \raisebox{-.5em}{\includegraphics[height=1.5em]{figures/icons/swipe-lr-finger.png}} and says ``\textit{region}."
By detecting the swipe gesture and the attribute \textit{Region}, the system infers that the user wants to arrange points horizontally by \textit{Region} and creates a unit column chart.


\subsubsection{Handling Ambiguity and Failure (\orgdc{DG4})}

To highlight ambiguity in commands, DataBreeze presents ambiguity widgets~\cite{gao2015datatone}. They appear in the feedback row and allow users to refine ambiguous values in the input command using pen/touch.
With the range of explicit, follow-up, multimodal commands and their possible phrasing variations (\orgdc{DG2}), interpretation errors are practically bound to occur during NL interaction.
\edit{When input commands are unintelligible (e.g., ``\textit{Apply a legion shelter}" instead of ``\textit{Apply a region filter}") due to speech recognition errors or beyond the scope of supported commands (e.g., ``\textit{Plot colleges geographically}"), DataBreeze notifies users about this failure, asking them to try a different command.}

\begin{figure}[t!]
    \centering
    \includegraphics[width=\linewidth]{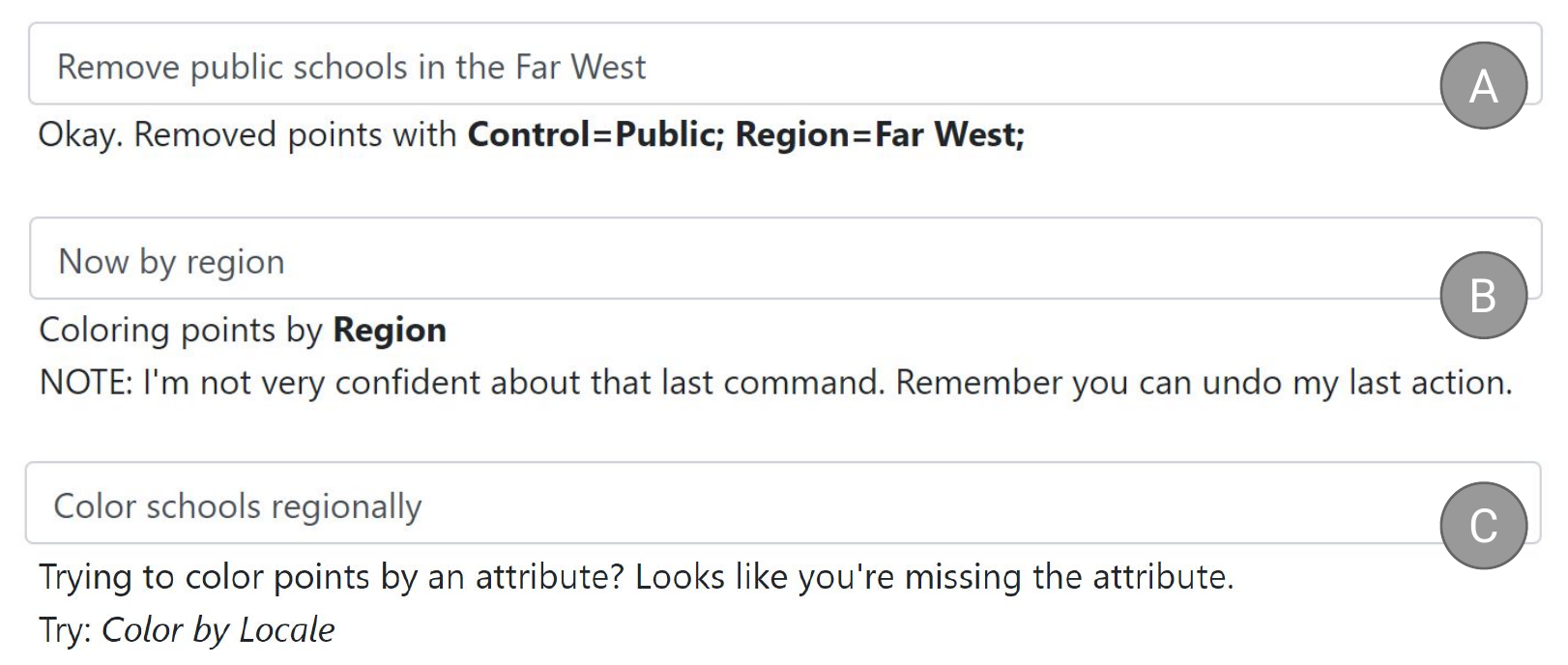}
    \caption{Feedback messages shown after (A) successfully executing a command, (B) executing a follow-up command, and (C) partially interpreting a command.}
    \label{fig:feedback-widgets}
\end{figure}

\edit{
A key difference between DataBreeze and existing visualization NLIs lies in how the system handles partially complete commands.
Current systems either only notify users about failure~\cite{hoque2018applying,srinivasan2018orko}, apply system defaults~\cite{hoque2018applying,srinivasan2018orko}, or list all possible operations or values to choose from~\cite{srinivasan2018orko}.
Instead, utilizing this failure as a teaching opportunity, DataBreeze performs an additional processing step and checks the command for partial phrasings or keywords that might map to potential operations.
If it finds a match, the system generates an explanation along with an exemplary command that could help the user learn the correct phrasing.
An example of this is shown in Figure~\ref{fig:feedback-widgets}C where the system suggests the example command \textit{Color by Locale} when it is unable to interpret the user command ``\textit{Color schools regionally}."
In this case, DataBreeze was able to determine the operation using the keywords `\textit{Color by}' but was unable to detect the attribute to color by (and thus randomly chose \textit{Locale} as an example: if there are multiple operations a partial command may map to, the system selects one at random).
By notifying users about its actions and providing suggestions and explanations when commands fail, DataBreeze attempts to reduce the ``black-box" effect of NL interaction~\cite{setlur2016eviza,srinivasan2017natural}.
}

\subsubsection{Command Feedback and Discovery (\orgdc{DG4})}

When DataBreeze processes a spoken command, it updates the feedback row to summarize the actions it performed in response to the command.
If the user command is successfully processed, the system states the operation along with target or parameter values (e.g., Figure~\ref{fig:feedback-widgets}A).
However, for follow-up commands, because the system infers user intent based on previous commands, there is a higher chance of error.
To highlight this, in addition to the action performed, the system feedback reminds users about the undo feature \edit{that allows reverting the most recent action} (Figure~\ref{fig:feedback-widgets}B).

\begin{figure}[t!]
    \centering
    \vspace{3mm}
    \includegraphics[width=.6\linewidth]{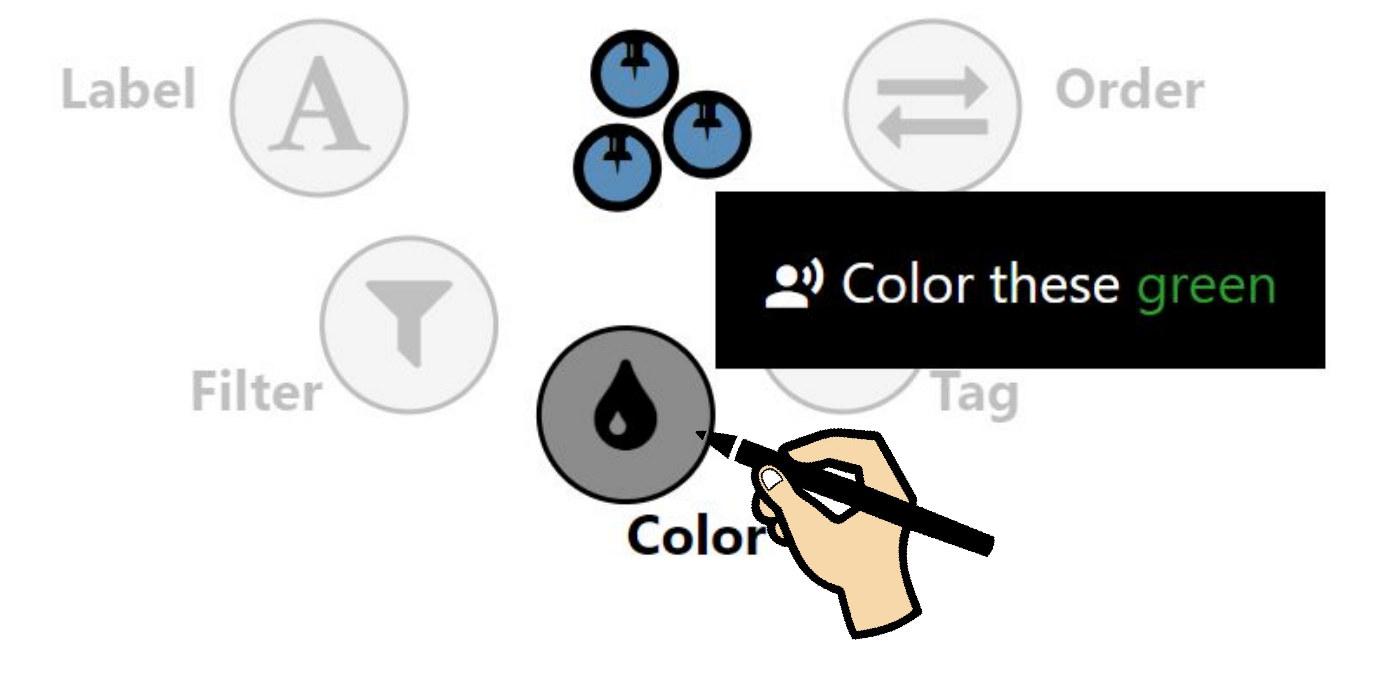}
    \vspace{-2mm}
    \caption{A sample deictic speech command for coloring selected points is shown on long pressing a context menu option.}
    \label{fig:voicehints-example}
\end{figure}

To preemptively make users aware of possible speech commands and phrasings, DataBreeze employs an adaptive command discovery approach~\cite{furqan2017learnability,srinivasan2019discovering}, suggesting commands through tooltips when users long press on context menu options (Figure~\ref{fig:voicehints-example}).
DataBreeze also suggests commands post-hoc once an operation has been performed using pen/touch.
To deliver these suggestions unobtrusively, the system displays them above the canvas in the feedback row.
For example, if the user removes labels for all points on the canvas using the context menu, the feedback row displays the message `{{\faLightbulbO}} To remove all labels, you could also \raisebox{-.15em}{\includegraphics[height=1em]{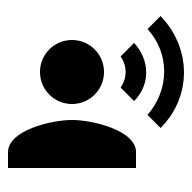}} \textit{Clear all labels}.'
Users can turn off the system suggestions by tapping the {\small{\faBell}} icon at the top right corner (Figure~\ref{fig:scenario-end}).

\subsection{System Implementation and Architecture Overview}

\begin{figure}[t!]
    \centering
    \includegraphics[width=\linewidth]{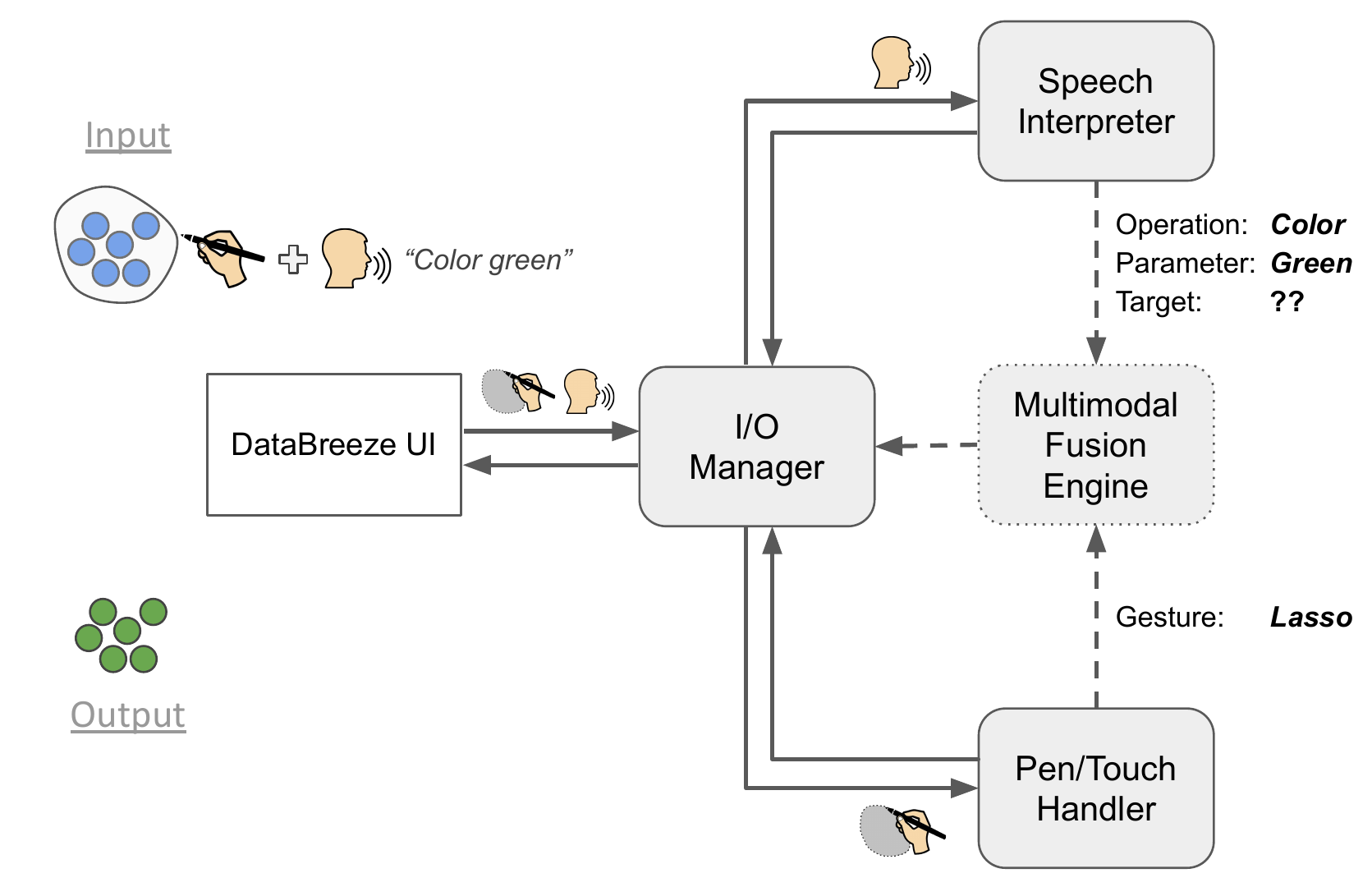}
    \caption{\added{DataBreeze System architecture highlighting the flow of information between different components for the exemplary interaction of coloring points by selecting them and saying ``\textit{Color green}."}}
    \label{fig:system-architecture}
\end{figure}

DataBreeze is implemented as a web-based application and supports data files with numerical and categorical attributes in the CSV format.
The visualization is rendered using D3.js~\cite{bostock2011d3}.
All pen/touch inputs are collected as standard JavaScript events and processed by custom event handlers.
DataBreeze uses the HTML5 webkit speech recognition for translating speech-to-text.
\added{At the start of each session, we train the recognizer with the data attributes and values from the input dataset as well as a list of system keywords (e.g., remove, summarize). While it still detects arbitrary speech, this training helps improve the recognition accuracy for the most relevant keywords in user commands.}
Translated commands are processed using a custom interpreter implemented in JavaScript.

Figure~\ref{fig:system-architecture} gives an overview of the system architecture.
At a high-level, all user input is collected by the \textit{I/O Manager} which then propagates it to the \textit{Speech Interpreter} or the \textit{Pen/Touch Handler} depending on the input type.
These components individually process the input and send the response with the required changes back to the I/O Manager which updates the DataBreeze interface.
For multimodal commands where these components are individually unable to process the input (e.g., the speech interpreter does not detect the required parameters to execute an operation), the extracted input details are passed to the \textit{Multimodal Fusion Engine}.
This component then combines the information from both input streams to determine the required system action and passes it back as a response to the I/O Manager.

\added{To combine different input streams, the fusion engine uses a predefined set of mappings between operations (e.g., coloring, assigning axes), their relevant parameters (e.g., color names, data attributes), and targets (e.g.~data points). For the example shown in Figure~\ref{fig:system-architecture}, the user selects a set of nodes \raisebox{-.5em}{\includegraphics[height=1.5em]{figures/icons/lasso-pen-custom.png}} with the pen (which implicitly triggers speech) and says ``\textit{Color green}." Using the keywords in the input command, the \textit{Speech Interpreter} identifies that the operation being referred to is \textit{Color} and the color to be used is \textit{green} (the operation parameter). However, because the command does not specify any criteria to select points (missing target for the \textit{Color} operation), the \textit{Fusion Engine} also considers the input from the \textit{Pen/Touch Handler}. Detecting that a selection (lasso) was performed, DataBreeze infers the target points by considering the active selection state of points on the canvas. Upon detecting the selected points, the \textit{Fusion Engine} combines this information with the output from the \textit{Speech Interpreter} and updates the color of the selected points.}

\edit{Compared to previous speech-based multimodal visualization systems like Orko~\cite{srinivasan2018orko} which perform the fusion temporally, the fusion engine in DataBreeze operates based on the semantics of the system state (e.g., selections, active filters) and the intended operation (e.g., creating axes, moving).
This allows using speech and pen/touch simultaneously or sequentially (starting with either modality).
For example, one can point \raisebox{-.5em}{\includegraphics[height=1.5em]{figures/icons/point-finger.png}} to a location on the canvas while speaking a command, or point and lift the finger to think about a command and then say it, both resulting in the same action.
DataBreeze also preserves gestures as part of the context objects across multimodal commands.
For instance, if one swiped \raisebox{-.35em}{\includegraphics[height=1.5em]{figures/icons/swipe-lr-finger.png}} and said ``\textit{Region}," DataBreeze will set the axis to \textit{Region}.
Now if the user issues ``\textit{Locale}" as the next command, the fusion engine preserves the swipe gesture in memory and repeats the axis specification operation but with the new \textit{Locale} attribute.}

\section{Preliminary User Study}

We conducted a preliminary user study to gauge people's reactions to DataBreeze, and more specifically to observe: (1) if and when people switch between systematically bound and manually customized views during data exploration and (2) the use of different modalities during data exploration with flexible unit visualizations.

\subsection{Participants and Experimental Setup}
We recruited six participants (P1-P6; four females, two males), aged 23 to 28.
All participants were university students and indicated their field of study as computer science (P1), HCI (P2, P3, P5), visual art (P4), and cognitive science (P6).
In terms of prior visualization experience, two participants (P1, P2) stated they had minimal experience working with visualization tools, three (P3, P4, P5) said they had worked with visualization tools on multiple occasions but not on a regular basis, and one participant (P6) said she was a frequent Tableau user.
All participants\added{~were native English speakers and} rated themselves as being moderately to highly comfortable working with touch-, pen-, and speech-based systems (e.g., on their iPads or Siri/Alexa).
Participants interacted with DataBreeze on Google's Chrome browser on an 84" Microsoft Surface Hub set to a 3840x2160 resolution \added{with an external microphone to capture voice commands (as illustrated in Figure~\ref{fig:teaser})}.
\edit{All sessions were audio and video recorded.}

\subsection{Procedure}
Participants were first given a brief introduction to DataBreeze including the interface components, the interactions they could perform with pen/touch, and how they could invoke speech (5 min).
To avoid biasing participants towards interacting in a particular way, we neither showed examples of speech commands nor gave them an exhaustive list of the available operations.
Next, as ``training,'' we directed participants to freely interact with DataBreeze so they would be comfortable with the different interactions (10 min).
We used a dataset about cars for the introduction and training phase.
Participants were free to ask any questions they had regarding the interactions or system behavior during the training phase.

Participants then performed an open-ended task with the colleges dataset (with 500 U.S.~colleges) in which they were asked to explore the data to produce a list of colleges for their younger siblings to apply to (15-30 min).
\edit{Participants were free to 
leverage their external knowledge of U.S.~colleges as the shortlisting criteria.}
We did not ask participants to think-aloud because it could result in unintended recognition (due to the implicit speech triggering techniques) and also interrupt their workflow.
We did, however, ask participants to take screenshots whenever they felt they identified a view that they would like to share with their family members (e.g., for discussion).
Furthermore, we told participants to let us know whenever the system did not behave as they expected.

The study session ended with a debriefing in which we asked participants to provide feedback on their experience working with the system (5-15 min).
We asked a set of standard questions across participants, but also seeded these interviews with our observations during the session (e.g., always using a specific modality for an operation).
Overall, sessions lasted between 40-60 minutes and participants were compensated with a \$20 Amazon Gift Card for their time.

\subsection{Results and Observations}

All participants completed the task, identifying at least ten colleges.
Four participants identified multiple groups of colleges based on different criteria (P1, P3, \& P5: two groups, P6: three groups).
After identifying each group, these participants took a screenshot and reset the tool to start over. This resulted in 11 shortlisted groups of colleges across the six participants.
Below, we highlight key observations from the study, focusing on the participants' data exploration patterns afforded by the combination of flexible unit visualizations and multimodal interaction.

\subsubsection{Visual Data Exploration Patterns}

We observed three high-level patterns that participants employed while exploring data with DataBreeze.

The most common pattern was participants starting with a systematically bound view (\textbf{SB}), switching to a manually customized view (\textbf{MC}), but later switching back-and-forth between the two views one or more times (\textbf{SB$\rightarrow$MC$\leftrightarrow$SB}).
We observed this pattern during five (out of 11) shortlists identified across four sessions (P1, P2, P4, P6).
For example, P4 started with a scatterplot of \textit{Average Cost} and \textit{Median Earnings}.
As he inspected colleges, he switched to a customized view where he removed points from the scatterplot and spatially categorized them into two virtual bins of ``General" and ``Field Specific" schools (\textbf{SB$\rightarrow$MC}).
Once he had narrowed down on a subset of colleges and refined his virtual bins by creating smaller, local scatterplots within the bins.
He then specified a global X-axis based on \textit{Region} (\textbf{MC$\rightarrow$SB}), directly manipulating the resulting view to order points in a custom manner within each region (\textbf{SB$\rightarrow$MC}).
This seamless transition between systematically bound and manually customized views was enabled by DataBreeze's support for global and local operations (\newdc{DG5}) as well as multimodal interaction (e.g., selecting a set of points and using swipe-and-speak to define a local axis, or issuing the same command without selecting points to create a global axis).

The second common pattern involved participants starting with a systematically bound view, then switching to a manually customized view and resorting to customized views until the end of their exploration (\textbf{SB$\rightarrow$MC}).
We observed this pattern during four shortlists across four sessions (P1, P3, P5, P6).
For example, P5 created a scatterplot visualizing \textit{Admission Rate} and \textit{Population}, coloring points by \textit{Locale}.
From this scatterplot, she selected a set of colleges with lower \textit{Admission Rate}s and ordered them by \textit{Average Cost}.
She continued to work with this subset of points creating new scatterplots with other attributes.
However, while she worked with these points, she preserved her original scatterplot and would go back to it to explore a different set of points.
This illustrates an interesting example similar to the one in Figure~\ref{fig:flexibility-spectrum}B, where the initial systematically bound view and global mappings become a platform to facilitate more localized exploration.

The last (i.e., least common) pattern comprised of participants exploring data exclusively using systematically bound views (\textbf{SB}).
This pattern was employed during two shortlists, once each by P3 and P6.
They created a scatterplot and iteratively refined it by filtering points or modifying visual encodings until they identified a group of points that were interesting to them.
This exploration strategy strongly aligns with Shneiderman's information seeking mantra~\cite{shneiderman1996eyes} and is largely supported by shelf configuration and control panel-based visualization tools today.



\subsubsection{Multimodal Interaction Usage and Feedback}

Participants performed a total of 164 key operations (e.g., assign X/Y axes, change color, filter) across all sessions.
\added{As shown in Table~\ref{tbl:interactions-summary},} 37 (22\%)\added{~of these interactions} were performed using speech alone (e.g., ``\textit{Color by region}"), 43 (27\%) were performed using only pen/touch and context menus (e.g., selecting points and using the context menu to remove them), and 84 (51\%) involved a combination of pen/touch and speech input (e.g., select-and-speak, swipe-and-speak).
\added{Besides the 121 speech commands that were correctly recognized (37 speech-only + 84 multimodal), there were seven misrecognized commands (0-3 commands per session).}
Overall, regardless of their exploration strategy or pattern, all participants leveraged both touch/pen and speech (either unimodally or multimodally) to interact with the system.

Participants preferred speech to perform operations globally (e.g., color by attribute) but context menus to perform operations locally (e.g., coloring specific points red).
When we asked participants about their choice to use context menus over select-and-speak (e.g.,\raisebox{-.5em}{\includegraphics[height=1.5em]{figures/icons/lasso-pen-custom.png}} + ``\textit{Remove}"), participants said they felt a stronger sense of control with menus especially because local operations involved deletions or making fine-tuned changes that were better suited for dynamic querying~\cite{shneiderman1994dynamic} afforded by menus (e.g., changing color of points using a color picker).

In general, participant feedback suggested that their choice of interaction was primarily based on reducing the time to perform an operation.
For instance, regardless of whether they were creating an axis for a subset of points (local operation) or for all points on the canvas (global operation), all participants used swipe-and-speak (e.g.,\raisebox{-.5em}{\includegraphics[height=1.5em]{figures/icons/lasso-pen-custom.png}} + ``\textit{Admission Rate}") attributing this to being both natural and fast.
P4, for instance, said ``\textit{especially because the system was already listening to me when I swiped, it made more sense to say the attribute name than go to a menu and choose an attribute}."
This comment also highlights the importance of subtle aspects such as triggering techniques for speech commands (\newdc{DG7}) during the design of multimodal systems.
\remove{The consistent use of multiple modalities across participants also further supports ongoing work on NL-based visualization systems.
Specifically, similar to prior work~\cite{setlur2016eviza,srinivasan2018orko}, our observations suggest that given the choice, people may prefer interacting with visualizations using NL over DM or control panels depending on the operation they are performing.}

\begin{table}[t!]
    \caption{\added{Summary of participants' interactions with DataBreeze.
    Cells show the number of occurrences of an interaction (rows) for each participant (columns).
    Cells are colored column-wise from white (no interaction) to dark blue (most frequent interaction) for each participant.}}
    \centering
    \includegraphics[width=\linewidth]{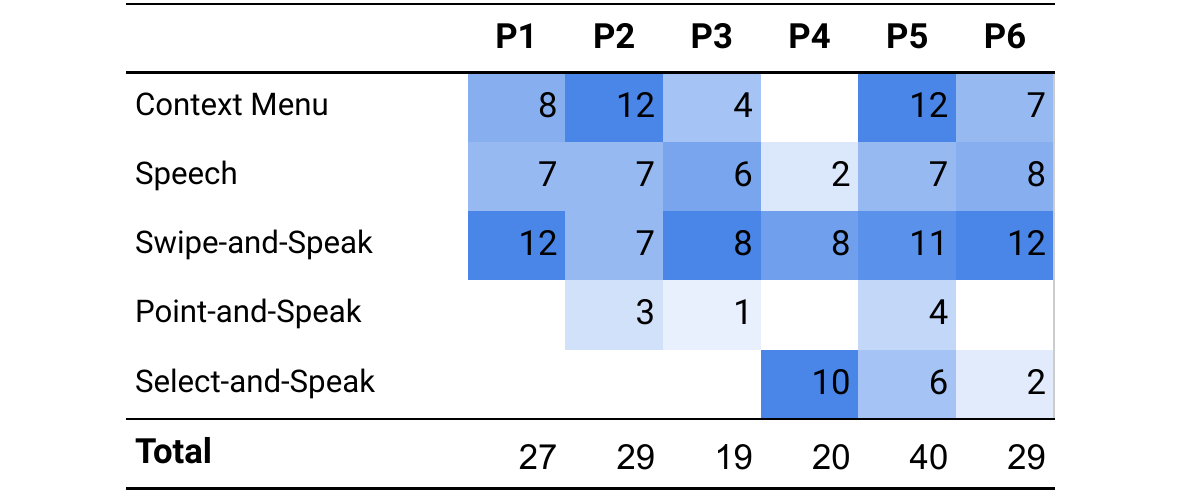}
    \vspace{-4mm}
\label{tbl:interactions-summary}
\end{table}

\section{Discussion and Future Work}

The iterative design process and the preliminary user study helped us identify \remove{some open design questions and}\edit{a number of key takeaways, raising additional design questions and highlighting} avenues for future work. We discuss these points below.

\subsection{Interweaving Interaction and \edit{Flexible} Representation Techniques to Design Novel Tools and Experiences}
DataBreeze's flexibility stems from the fact that it interweaves multimodal interaction combining DM and NL with a particular visual representation (flexible unit visualizations) that provides suitable affordances for that style of interaction.
For instance, talking about the ability to drag points out of a scatterplot during her interview, P5 said ``\textit{I really like that flexibility. Being able to drag and drop points where you want mimics physical interaction like you would with documents on our table.}"
This exemplifies how, in this case, the \textit{directness of manipulation}~\cite{hutchins1985direct} afforded by pen and touch naturally lent itself to participants expecting flexibility not supported in a common visual representation (dragging points in a scatterplot).

Another participant (P4) with a design background compared his experience using DataBreeze to that of creating a mood board with a physical art board.
In this context, he stated ``\textit{The system was great for that [exploring data similar to using an art board]. I could just quickly drag and pull things to create groups and categories that made sense in my head.}"
Later, referring to his frequent usage of operations like coloring or filtering specific points through select-and-speech actions, he said, ``\textit{sometimes voice can be more of a novelty than a tool but in this case, it felt definitely like a tool. Where, if I had to otherwise keep going to some buttons and pressing them I would probably have used them less.}"
Once again, such comments collectively suggest that the combination of the affordances of the visual representation and input/interaction techniques resulted in DataBreeze supporting a workflow that is closer to how people interact with objects in the real-world (in P4's case, with an art board).

Although DataBreeze is just one example of how this may be accomplished, it illustrates the potential of novel tools and experiences that can emerge by \textit{interweaving interaction techniques and \edit{flexible} representations}, adjusting each as necessary to enable a seamless user experience.
While the two themes of interaction and representation have individually received considerable attention in visualization research today, far fewer examples have explored new tools and experiences that emerge from their synergistic integration (e.g.,~\cite{claessen2011flexible,rzeszotarski2014kinetica,cordeil2017imaxes,le2019dynamic}).
With interactive devices and interfaces supporting alternative forms of input becoming a more common platform for visualizations, a \remove{unique}\edit{compelling} research opportunity lies in exploring visualization tools and user experiences stemming from the combination of naturalistic input/interaction techniques and more flexible visual representations.

\subsection{Leveraging Complementarity-based Multimodal Interaction}

A \remove{unique aspect}\edit{key feature} of DataBreeze \edit{differentiating it from previous NL-based multimodal visualization systems} is its increased support for \textit{complementarity-based} multimodal interaction~\cite{martin1998tycoon}, where individual modalities are used to acquire different chunks of information which are then merged to enable a more sophisticated operation.
For example, in a swipe-and-speak action, the operation of specifying the location and direction of an axis is specified through a pen/touch gesture whereas the designation of which data attribute to place on that axis is done through speech.
Participants frequently performed such interactions and commented on them favorably.
This was also reflected by the high number (84/164) of multimodal interactions during the user study.
In fact, even participants who were initially hesitant about multimodal interaction (P3, P4, P6) quickly adapted to such interactions.
For instance, P6 said ``\textit{Gestures typically in my mind aren't combined with voice commands [...] but once I got used to it, it was great and saved a lot of time.}"
Referring to the swipe-and-speak action, P5 specifically noted that she found complementarity-based multimodal interaction to be most effective when there was a strong direct mapping between an operation and the input modality.
As also highlighted earlier (\newdc{DG7}), an inherent consideration for supporting complementarity-based multimodal interaction was implicitly triggering speech input at the right time so that participants can more seamlessly integrate their actions across modalities.
Along these lines, five out of the six participants (except P2) commented favorably on the implicit triggering techniques stating it made interacting with the system both fast and more natural.

This feedback suggests that if employed for the right operations, complementarity-based multimodal interaction can be a valuable feature in visualization systems by supporting a more fluid, integrated interaction experience~\cite{walny2012understanding}, in turn, helping people preserve their workflow.
Furthermore, spoken commands in complementarity-based interactions are typically shorter and more focused (e.g., including only parameter values like attribute names or operation specific keywords like color, size, etc.).
In addition to being easier for users to speak, from a system standpoint, these commands are generally easier to interpret~\cite{vo1993multimodal}.
With these potential user- and system-centered benefits in mind, an open area for future exploration lies in identifying operations and tasks that are best suited for complementarity-based multimodal interaction, as well as the appropriate interface support to mediate such interactions.

\subsection{Complementing Open-ended Exploration with Targeted Question Answering}
With DataBreeze, we largely focused on examining how interweaving flexible unit visualizations with multimodal interaction can aid visual data exploration.
Because one may want to get data value summaries for subsets of points, DataBreeze also supports generating visual summaries through histograms. 
However, participants additionally wanted to get more targeted information or details during their exploration.
These needs for details would often stem from participants' observations based on the view.
For example, after identifying a set of interesting schools, P1 asked ``\textit{What is the average Expenditure for these colleges?}"
Similarly, noticing the high cost of schools in \textit{New England}, P3 asked ``\textit{Which is the most expensive school in the Far West?}"

While these questions could be answered by stating the requested values and/or data cases, the fact that they were asked in the context of a specific visual state makes them semantically meaningful and interesting.
Addressing such scenarios, one promising direction for future work is developing NL interpreters that are capable of answering explicit (e.g., ``\textit{Which is the most expensive public school?}") and multimodal (e.g.,\raisebox{-.5em}{\includegraphics[height=1.5em]{figures/icons/lasso-pen-custom.png}} + ``\textit{What is the average SAT score for applying to these schools?}") data-driven questions.
In addition to computing the response, another challenge is designing appropriate feedback mechanisms to present these responses in the visual context that the question was asked (e.g., showing a temporary annotation to indicate the requested average value).


\subsection{Providing Appropriate Default Behaviors}

\textbf{Implicit data grouping.}
Both during the design phase and the preliminary study, participants often wanted to operate on spatially co-located groups of points without explicitly selecting points.
Consider the scene in Figure~\ref{fig:scenario-1}E. After ordering the \textit{Far West} schools, currently, one cannot issue a command like ``\textit{Reorder all groups}" to order schools from other regions.
Instead, one has to individually select points from other regions and order them.
From a system standpoint, supporting a command like ``\textit{Reorder all groups}" would require implicitly determining what the groups are.
Such implicit grouping imposes additional questions regarding the implementation and interface design, however.
For instance, should points that are selected or moved together be considered a group?
Or should the system infer groups based on the X/Y co-ordinates of points?
Should these implicitly determined groups be added as targets for follow-up commands?
Furthermore, how should the system visually highlight implicit groups?
As a first step, we allow users to select points and tag them to form ``virtual" groups that can be accessed using a common tag.
However, incorporating more nuanced methods to determine groups and perform actions on them is an immediate area for potential improvement in DataBreeze.

\bpstart{The default role of pen input}
In response to initial confusion regarding the unequal roles of pen and touch during the design phase, DataBreeze largely treats pen and touch interchangeably (i.e., both can be used to move or select points) (\newdc{DG6}).
Although DataBreeze also supports the use of pen as an inking tool (\newdc{DG8}), its support for drawing, annotations, and note-taking is limited compared to other visualization tools that focus on bimanual interaction combining pen and touch.
\edit{These systems have shown the value of using a pen as an inking or sketching tool for both visualization authoring (e.g.,~\cite{frisch2009investigating,frisch2010diagram,lee2013sketchstory,xia2018dataink}) and sensemaking (e.g.,~\cite{romat2019activeink,kim2019inking}).}
Based on these systems, one approach is to default the pen's primary operation to inking (mimicking its function in the real-world).
However, if the pen is used to ink, operations such as selections and specifying axes may need to rely on other modalities.
Correspondingly, one open area for future research is to investigate how changing the primary role of the pen affects the interaction and interface design of a multimodal visualization system like DataBreeze that supports touch and speech as alternative modalities.

\subsection{Improving System Feedback \edit{and Error Recovery}}

An important aspect of the interface design was to provide appropriate feedback in response to spoken commands (\orgdc{DG4}).
Correspondingly, we reserved an exclusive region in the interface for feedback directly under the speech input box (Figure~\ref{fig:scenario-end}B).
Although the system presented different types of feedback, even suggesting corrections when possible, during the user study, participants often failed to notice the feedback.
This was particularly problematic when there were command phrasing related errors.
Since participants did not see the feedback, they ignored the system's phrasing suggestions and instead hyperarticulated their initial commands~\cite{myers2018patterns}, resulting in the same error.
Hence, an open area for improvement in DataBreeze is to examine alternative feedback techniques that are more noticeable yet unobtrusive to the user's workflow.

\edit{A related point to feedback is error recovery.
Similar to other speech-based mutlimodal systems (e.g.,~\cite{vo1996building,tse2006enabling}), DataBreeze allows users to undo the most recent voice command.
Going forward, it is important to implement a more complete undo stack, tackling associated challenges in doing so (e.g., managing scope~\cite{abowd1992giving}, handling errors in undo command utterances~\cite{halverson1999beauty}).
With the flexibility of creating custom views and making global versus local changes, giving users the ability to backtrack multiple steps would further enhance the overall usability and user experience.}
\section{Limitations}

Our observations in this paper are based on testing DataBreeze on an 84" vertical display with up to about 1,200 data points.
While the described interactions may work across different display sizes, the fact that each mark needs to be large enough (to be interacted with a finger) raises concerns from a scalability standpoint.
Thus, an important next step is to investigate the scalability of the proposed approach by testing it on different displays (e.g., tablets, touch-enabled PCs) with varying dataset sizes.
\remove{The preliminary user study with six participants performing an open-ended task helped us assess the overall usage and feasibility of the proposed approach.
However, the observations and subjective feedback cannot substitute for a more detailed qualitative study that investigates specific interactions and the use of different modalities over the course of multiple tasks or sessions.
Thus, an important next step is to leverage DataBreeze as a platform to further investigate multimodal interaction with visualizations, understanding the accompanying cognitive and practical benefits and challenges.
Although our study participants subjectively stated that they enjoyed using the tool and felt engaged in the exploration, we did not formally measure or track these metrics.
Thus, another important factor to consider as part of future studies will also be to assess the impact of multimodal interaction on user experience-focused metrics~\cite{saket2016beyond} such as engagement and enjoyment.}

\edit{Our goal in designing DataBreeze was to explore if and how multimodal interaction along with flexible unit visualizations can facilitate free-form data exploration.
The preliminary user study with six participants performing an open-ended task helped us assess the overall usage and feasibility of the proposed approach.
However, our study does not identify potential benefits of the proposed approach compared to conventional systematically-bound visualizations such as maps and scatterplots, and the possible cognitive challenges associated with creating and interacting with flexible unit visualizations.
Furthermore, while our study helps validate system usability, it cannot replace a more detailed qualitative study that investigates specific interactions and the use of manually customized views over the course of multiple tasks or sessions.
Therefore, it is important to conduct follow-up studies with systems like DataBreeze, investigating multimodal interaction with flexible representations to better understand the accompanying cognitive, physical, and analytic benefits and challenges.
}
\section{Conclusion}

\edit{Through the design and implementation of DataBreeze, we exemplify how interweaving DM- and NL-based multimodal interaction with flexible unit visualizations enables a novel data exploration experience.
Specifically, we discussed how allowing people to multimodally create and interact with both systematically bound and manually customized views empowered them to freely explore the data and update the visualization to reflect their mental models.}

\edit{Our observations coupled with participant feedback during the design sessions and a user study collectively highlight promising areas for future research including:
(1) leveraging the complementary nature of pen, touch, and speech to enable a fluid interaction experience during data exploration,
(2) facilitating targeted question answering during open-ended visual data exploration,
and (3) supporting feedback and error recovery mechanisms for speech input that are noticeable yet unobtrusive. 
Although DataBreeze is only one example, it highlights exciting research opportunities and challenges in developing multimodal systems facilitating more fluid and natural interaction with data.
We hope this work inspires the design and development of a new generation of post-WIMP interfaces for data visualization that empower our human perceptual, cognitive, and manipulative abilities.}

\ifCLASSOPTIONcompsoc
  \section*{Acknowledgments}
\else
  \section*{Acknowledgment}
\fi

This work was supported in part by the National Science Foundation grant IIS-1717111.

\ifCLASSOPTIONcaptionsoff
  \newpage
\fi



\bibliographystyle{IEEEtran}
\bibliography{databreeze.bib}
%



%

\begin{IEEEbiography}[\vspace{-2.5em}{\includegraphics[width=1in,height=1.25in,clip,keepaspectratio]{{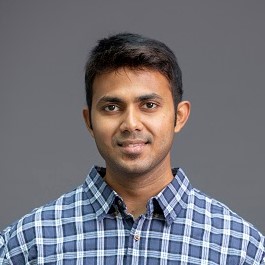}}}]{Arjun Srinivasan}
is a Ph.D.~candidate in Computer Science at the Georgia Institute of Technology. His current research focuses on the design of intelligent and expressive visualization tools that combine multimodal input (e.g., speech and touch) and mixed-initiative interface techniques for human-data interaction.
\end{IEEEbiography}

\begin{IEEEbiography}[\vspace{-1.5em}{\includegraphics[width=1in,height=1.25in,clip,keepaspectratio]{{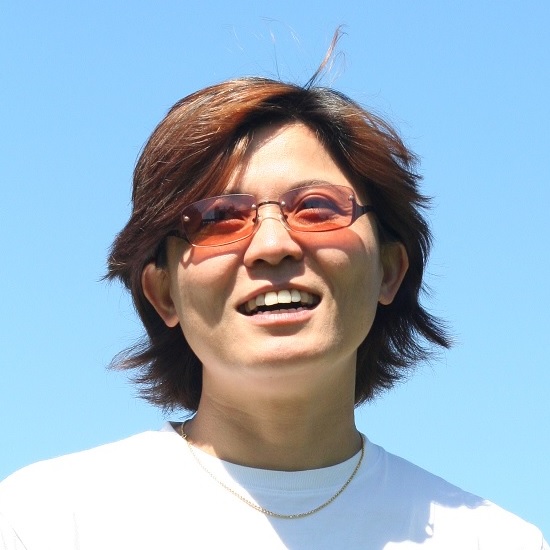}}}]{Bongshin Lee}
is a Sr. Principal Researcher at Microsoft Research. 
She explores innovative ways for people to create visualizations, interact with their data, and share data-driven stories visually. She has also been recently focusing on helping people explore the data about themselves and share meaningful insights by leveraging visualizations. She received her PhD in Computer Science from the University of Maryland at College Park in 2006.
\end{IEEEbiography}


\begin{IEEEbiography}[{\includegraphics[width=1in,height=1.25in,clip,keepaspectratio]{{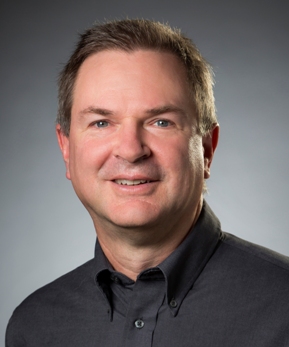}}}]{John Stasko}
is a Regents Professor in the School of Interactive Computing and the Director of the Information Interfaces Research Group at the Georgia Institute of Technology. His research is in the areas of information visualization and visual analytics, approaching each from a human-computer interaction perspective. John was named an ACM Distinguished Scientist in 2011 and an IEEE Fellow in 2014. He received his PhD in Computer Science at Brown University in 1989.
\end{IEEEbiography}




\end{document}